# High-Yield, 150-ps Polymer Scintillators via Synergistic Purcell and High-Z Enhancement

Xiaohe Zhou[1], Hiba H. Karakkal[1], Chenger Wang[1], Andrea Fratelli[1,3], Francesco Bruni[1,2], Matteo L. Zaffalon[1,2], Leonardo Poletti[4], Saptarshi Chakraborty[1], Emanuele Mazzola[5], Roberto Lorenzi[1], Francesco Carulli[1], Francesca Rossi[4], Francesco Meinardi[1], Luca Gironi[2,5] and Sergio Brovelli*[1,2]

[1] *Dipartimento di Scienza dei Materiali, Università degli Studi di Milano-Bicocca, Via R. Cozzi 55, 20125, Milano, Italy*

[2] *INFN - Sezione di Milano - Bicocca, Milano 20126 - Italy*

[3] *Nanochemistry, Istituto Italiano di Tecnologia, Via Morego 30, 16163, Genova, Italy*

[4] *IMEM-CNR, Parco Area delle Scienze 37/A - 43124 Parma, Italy*

[5] *Dipartimento di Fisica, Università degli Studi di Milano-Bicocca, Piazza della Scienza 3, 20126 Milan, Italy*

Email: sergio.brovelli-unimib.it

* Corresponding author

This work is dedicated to the memory of Gianluca Latini, colleague and friend.

**ABSTRACT**

**Developing plastic scintillators that simultaneously achieve high light yield (*LY*), ultrafast timing, and scalable processability remains a formidable challenge in radiation detection. Here, we report a solution-processable hybrid scintillator that integrates high-Z sensitization with Purcell-enhanced radiative recombination within a conjugated-polymer platform. By embedding $HfO_2$ nanoparticles and resonant Ag-$SiO_2$ plasmonic nanoantennas into a poly(9,9-dioctylfluorene) matrix, we exploit the intrinsic sub-nanosecond kinetics of conjugated polymers while overcoming their characteristic low stopping power. The $HfO_2$ nanocrystals amplify energy deposition and suppress interchain aggregation, preserving high photoluminescence quantum yields in the solid state. Simultaneously, the nanoantennas induce strong plasmon-exciton coupling, accelerating radiative decay to ~150 ps. This synergistic architecture yields *LY*=6.1 photons/keV, comparable to standard commercial plastics with nanosecond lifetimes, yet over an order of magnitude higher than current sub-nanosecond state-of-the-art materials. This performance represents one of the most favourable combinations of efficiency and timing resolution reported to date. Our versatile strategy provides a framework for next-generation organic scintillators, merging record-breaking temporal response with high efficiency for advanced radiation-sensing architectures.**

Ionizing radiation detection is a cornerstone of technologies ranging from medical imaging and high-energy physics to security screening and industrial diagnostics[1]. Scintillators are among the most widely employed radiation converters, translating high-energy photons and particles into detectable optical signals. Inorganic crystalline scintillators such as NaI:Tl, CsI:Tl and LYSO currently set the benchmark for light yield (*LY*), defined as the number of emitted photons per unit deposited energy, and for energy resolution. However, their relatively slow scintillation decay, typically spanning tens to hundreds of nanoseconds, fundamentally limits their performance in fast-timing applications, including time-of-flight positron emission tomography and high-rate particle detection. In addition, their high production cost, brittleness and limited scalability hinder the development of large-area and mechanically compliant detectors[2]. Organic plastic scintillators provide a complementary platform that overcomes many of these limitations[3,4]. Their emission originates from fast electronic transitions in molecular fluorophores embedded within polymer matrices, enabling decay times in the nanosecond or sub-nanosecond (when quenched) regime together with low cost, mechanical flexibility and scalable fabrication. However, their intrinsically low atomic number severely restricts radiation stopping power and reduces detection efficiency, particularly for high-energy photons. This fundamental limitation is motivating the development of hybrid nanocomposite scintillators[5-12], in which high-Z nanoparticles (NPs) are incorporated into luminescent polymer hosts. In these systems, the inorganic phase enhances radiation absorption and generates cascades of secondary electrons that excite nearby organic emitters on sub-picosecond timescales, offering a route to increased LY while preserving the intrinsically fast response of plastic scintillators[13,14]. Reported examples include polystyrene or polyvinyl toluene matrices loaded with $Bi_2O_3$[15], $HfO_2$[8], or $Gd_2O_3$ NPs[8], as well as semiconductor quantum dots and perovskite nanocrystals (NCs) that act as sensitizers or fluorophores owing to their radiation hardness, tuneable bandgap and ultrafast multiexciton scintillation[5-7,16-21].

Despite these advances, achieving simultaneous optimization of scintillation speed and photon yield remains a major unresolved challenge. Even in state-of-the-art plastic scintillators (see **Figure 1a, Supplementary Table 1**), emission decay typically occurs over several nanoseconds, which is insufficient for emerging applications requiring extremely high-count rates or coincidence time resolutions approaching the tens of picoseconds regime. Existing strategies to accelerate scintillation predominantly rely on luminescence quenching, which reduces excited-state lifetimes below one nanosecond (700 ps in the fastest commercial materials) by introducing non-radiative relaxation pathways, inevitably driving the light yield down to a few hundred photons per MeV (orange line in **Figure 1a**). Consequently, sub-nanosecond emission remains largely inaccessible to high-photon-output materials. This limitation is increasingly motivating efforts in the emerging field of radiation quantum sensing[1,22-25], where collective phenomena are harnessed to achieve unprecedented speed and coherence in scintillation processes.

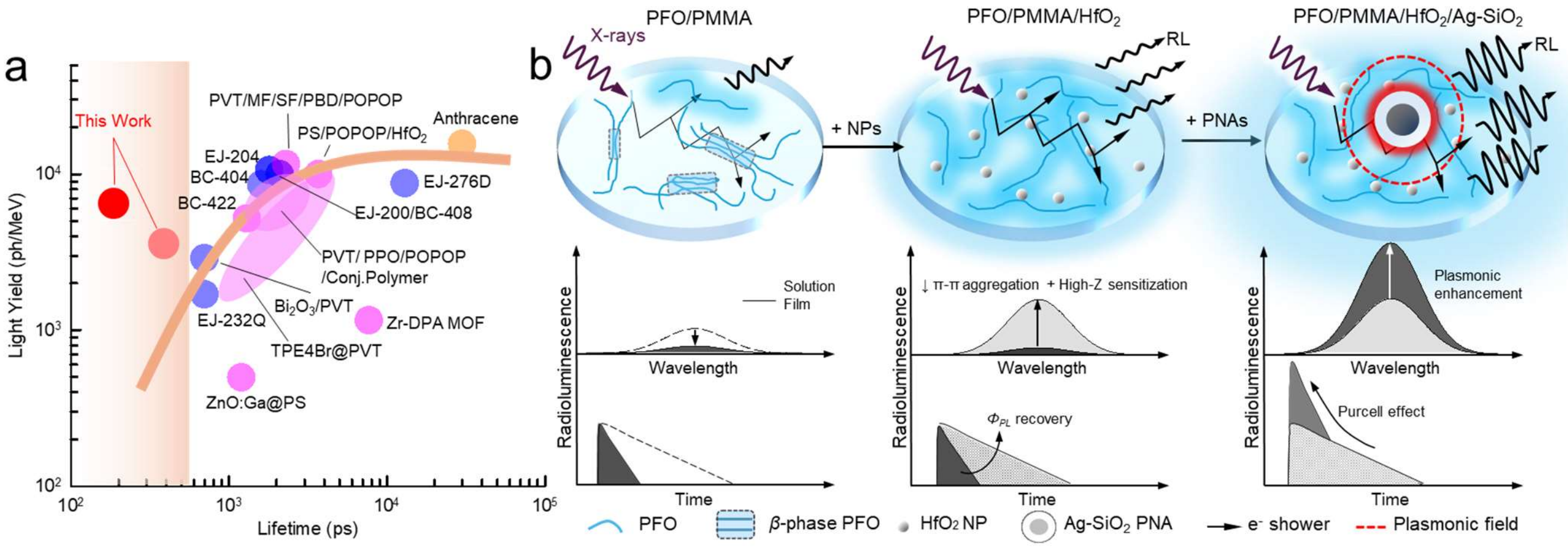

**Figure 1 | Time-efficiency trade-offs in plastic scintillators and concept of combined high-Z and plasmonic enhancement.** **a**, Light yield (*LY*) as a function of emission lifetime for commercial (blue symbols) and representative research (pink symbols) plastic scintillators, highlighting the intrinsic trade-off between photon output and temporal response [3,4,14,15,26-29]. The orange line indicates the reduction in LY observed in quenched, sub-nanosecond materials. **b**, Schematic of the proposed dual-enhancement strategy. *Left*: bare PFO/PMMA blend, where PFO forms both emissive glassy (α) and poorly emissive crystalline (β) phases; interchain aggregation accelerates emission non-radiatively at the cost of reduced fluorescence intensity. *Middle*: PFO/PMMA sensitized with $HfO_2$ NPs; $HfO_2$ prevents PFO aggregation, reducing non-radiative losses and enhancing $\Phi_{PL}$. Under ionizing radiation, high-Z NPs capture energy and release photoelectrons that excite PFO along their ionization track. *Right*: dual-enhanced PFO/PMMA with $HfO_2$ NPs and Ag–$SiO_2$ plasmonic nanoantennas (PNAs); plasmonic coupling intensifies the local electromagnetic field, boosting exciton generation and accelerating radiative scintillation dynamics.

Overcoming this efficiency-timing trade-off requires approaches that directly modify the intrinsic radiative dynamics of scintillation rather than relying on non-radiative deactivation processes. From a photonics perspective, the spontaneous emission rate of luminescent centres is not solely determined by their internal electronic structure but can be engineered through the local electromagnetic environment via control of the local density of optical states. The Purcell effect[30,31], originally formulated in the context of cavity quantum electrodynamics, provides a powerful framework for accelerating radiative recombination by coupling emitters to resonant photonic or plasmonic nanostructures[6,32,33]. In particular, plasmonic metallic nanoparticles enable strong electromagnetic confinement at deeply subwavelength scales (i.e. hot spots)[34], offering the possibility of enhancing radiative decay rates while preserving or even increasing emission efficiency. These advances have laid the foundation for the exploration of Purcell-enhanced scintillation using photonic crystal cavities or plasmonic coupling[35-41]. When combined with high-Z nanostructures that improve radiation absorption, such photonic engineering strategies open new pathways toward ultrafast scintillators that simultaneously deliver high *LY* and unprecedented sub-ns temporal resolution. Despite significant progress and encouraging theoretical results, a platform that simultaneously harnesses high-Z sensitization and Purcell enhancement to bridge the gap between competitive light yield and picosecond-scale timing resolution remains a missing link in radiation detection.

Beyond the advantages of coherent emitter-plasmon coupling, developing efficient and ultrafast sensitized scintillators requires organic emitters that outperform standard molecular fluorophores with nanosecond lifetimes[3,4]. Conjugated polymers are particularly attractive for this purpose due to their favourable optoelectronic properties and processing versatility. Their main advantage for ultrafast scintillation is an intrinsically high radiative decay rate, arising from exciton delocalization over several repeating units along the polymer backbone. This delocalization provides spectral tunability and yields a substantially increased oscillator strength, akin to the "giant oscillator strength" observed in traditional semiconductor systems.[27,42] As a result, these materials typically exhibit sub-nanosecond radiative lifetimes[3,27] combined with near-unity photoluminescence (PL) quantum yields ($\Phi_{PL}$). As we show below, when paired with appropriate high-Z sensitization strategies and Purcell-enhanced architectures, such properties allow access to very high RL speeds, aligning with current efforts to develop ultrafast plastic scintillators for high-luminosity collider experiments and for fast-timing medical imaging modalities such as positron emission tomography - in particular as high-density scintillator coatings or as metascintillators integrated with high-Z substrates such as LYSO or BGO[43-45]. Achieving conjugated-polymer-based Purcell scintillators, however, requires addressing a well-known limitation of these materials: their propensity to form π–π stacked aggregates in the solid state, whether as neat films or blends[46-48]. Such aggregation often induces dimeric or excimer species that quench molecular luminescence and exhibit diminished emission efficiencies, slower decay kinetics, and compromised spectral purity[49,50]. As we demonstrate here, this long-standing drawback is inherently mitigated when conjugated polymers are blended with NP sensitizers, which physically disrupt interchain stacking and thereby restore isolated-chain photophysics with near-unity emission yields in sensitized polymeric solids.

In this work, we introduce a fully solution-processable hybrid scintillator that integrates plasmonic near-field enhancement and high-Z sensitization within a conjugated-polymer matrix free from detrimental intermolecular aggregation, yielding ~150 ps radiative scintillation lifetimes and $LY$~6100 ph/MeV, which is comparable to commercial plastic scintillators with ns-long lifetimes and over one order of magnitude higher than ultrafast quenched plastic scintillators (e.g. EJ232Q, BC-422-5). Specifically, architecture combines poly(9,9-dioctylfluorene) (PFO) as the scintillating polymer, hafnium oxide ($HfO_2$) NPs as transparent high-Z sensitizers, and silica-coated silver nanoparticles functioning as plasmonic nanoantennas (PNAs) (**Figure 1b**). Thereby representing the first example of synergistic effect of high-Z sensitization and Purcell effect hybrid nanocomposite scintillator. Adding poly(methyl methacrylate) (PMMA) into the blend further maintains high scintillation intensity while reducing the required PFO fraction and overall material cost. The $HfO_2$ NPs not only enhance radiation stopping power without compromising optical transparency but also introduce mild structural disorder that suppresses π–π stacking and β-phase formation in PFO[51-53], thereby recovering the high emission yield characteristic of isolated chains in solution. PNPs, engineered with silica shells, exhibit spectral resonance with the PL and RL of PFO while preventing fluorescence quenching and enabling controlled plasmon-exciton coupling. The resulting PFO/$HfO_2$/Ag-$SiO_2$ composite unifies high-Z energy deposition and conversion with plasmonic modulation in a single scalable platform. Steady-state and time-resolved spectroscopies reveal a direct correlation between lifetime shortening, emission enhancement, and plasmonic modification. Overall, this streamlined integrated strategy establishes a clear experimental connection between plasmonic coupling and scintillation dynamics in soft-matter systems and provides quantitative design criteria for hybrid organic scintillators combining high $LY$, ultrafast response, and scalable manufacturability.

**Results and Discussion**

PFO was purchased from Ossila (cas#19456-48-5, $M_w$=2.6×$10^5$ g/mol, high-purity grade, molecular structure in **Figure 2a**) and used as received. Its optical response in dilute solution in chloroform is shown in **Figure 2b**, featuring the characteristic π–π* absorption band at 388 nm and a well-resolved vibronic PL progression with maxima at 417 nm (0–0), 448 nm (0–1), and 478 nm (0–2). The PL yield is $\Phi_{PL}$ = 87±8%, with essentially single exponential kinetics with lifetime $\tau_{CHCl_3}$~ 300 ps (See **Supporting Fig.S1**, *vide infra*, **Figure 3b**), confirming the suitability of PFO as ultrafast emitter. $HfO_2$ NPs were synthesized using a modified version of the procedure reported by Pei *et al.*[8]. Specifically, hafnium(IV) trifluoroacetate ($Hf(CF_3COO)_4$) was thermally decomposed at 330 °C under reflux in a high-boiling oleylamine (OAm) medium, yielding OAm-capped $HfO_2$ NPs whose surface was subsequently partially exchanged with bis(2-(methacryloyloxy)ethyl) phosphate (BMEP) (see **Supplementary Information** for details). The resulting NPs exhibit a quasi-spherical morphology with an average diameter of 4.2 ± 1.9 nm (**Figure 2d, e**). Thermogravimetric analysis (TGA) of the resurfaced $HfO_2$ NPs is presented in **Supplementary Fig. S2**, indicating an approximate 30 wt% mass loss attributed to surface-bound ligands. Owing to their wide electronic bandgap (~5.6 eV) and the effective suppression of surface defect states, which minimizes

Urbach-tail broadening, these $HfO_2$ NPs show no detectable sub-bandgap absorption in the UV-visible spectral region. As a result, they do not reabsorb PFO emission, confirming their suitability as optically transparent high-Z sensitizers.

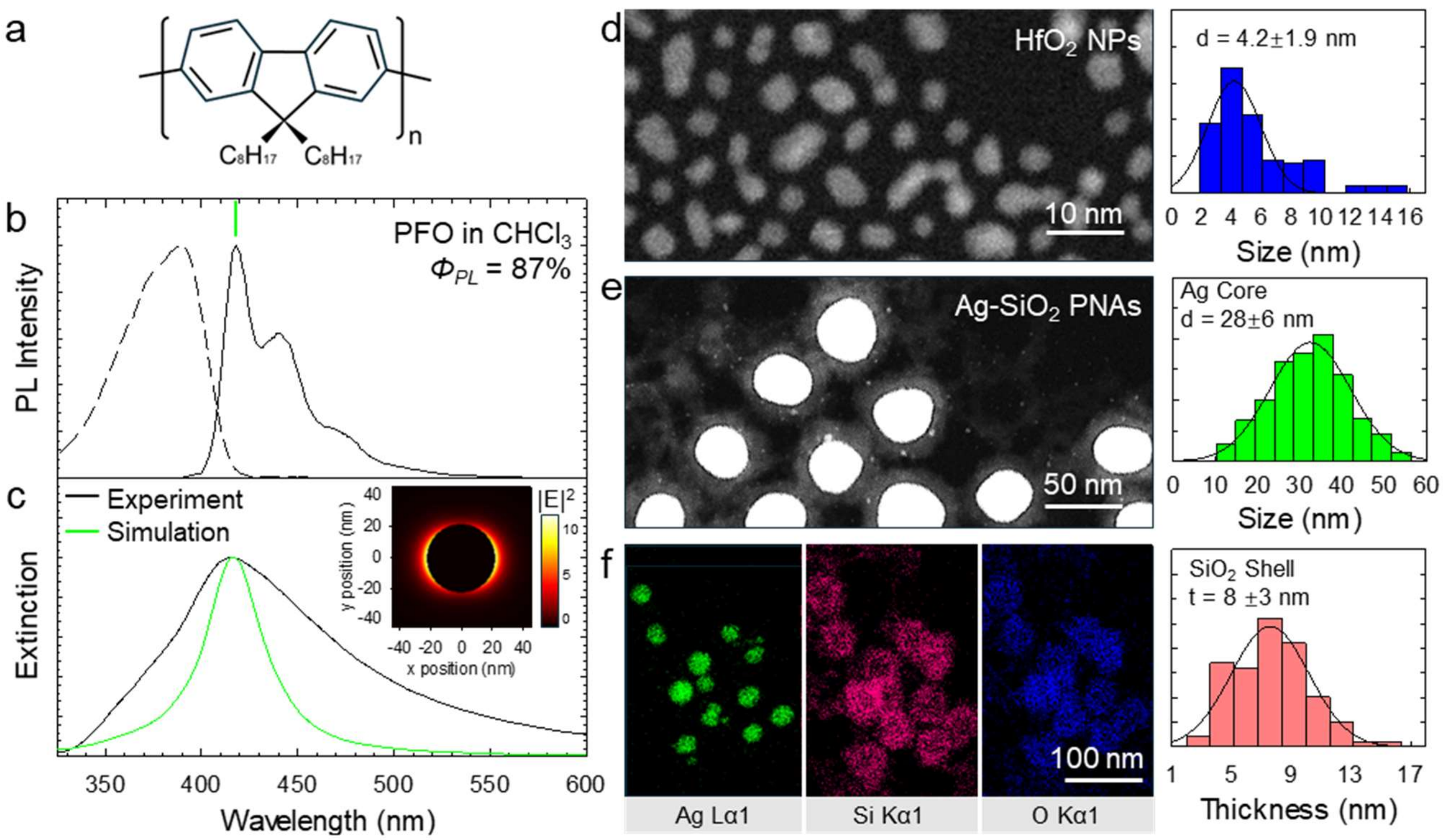

**Figure 2. Optical and morphological characterization of the blend constituents. a.** Chemical structure of poly(9,9-dioctylfluorene) (PFO). **b.** Normalized optical absorption (dashed line) and PL (black) spectra of PFO in chloroform, green mark on the top indicates the plasmonic resonance of the Ag-$SiO_2$ PNAs at 417 nm. **c.** Experimental extinction spectrum of the Ag-$SiO_2$ PNA (black) together with the simulated extinction spectrum (green) for a single nanoparticle embedded in a silica shell. The inset shows the calculated near-field intensity distribution at plasmon resonance wavelength. **d.** STEM-HAADF image and corresponding size-distribution histogram of $HfO_2$ NPs. **e.** STEM-HAADF image of core–shell Ag–$SiO_2$ PNAs and composite STEM-EDX elemental map **f.** STEM-EDX single element maps acquired from the same region as in **e**, confirming the core–shell architecture and chemical composition of the Ag–$SiO_2$ PNAs. Right panels correspond to size-distribution histograms of the Ag core and $SiO_2$ shell thickness.

Ag PNAs were engineered to exhibit a localized surface plasmon resonance (LSPR) matched to the main emission peak of PFO at 417 nm, enabling controlled plasmon-exciton coupling in the composite system. Based on a quasi-static (electrostatic) polarizability model for core–shell spheres, the Ag core radius was set to 14 nm. Following iterative optimization cycles, the $SiO_2$ shell was chosen at ~8 nm, which was the minimum thickness to suppress quenching of the PFO emission, ensure colloidal stability in polar solvents, prevent surface oxidation of the Ag cores and maintaining a controlled dielectric separation that enables reproducible plasmonic coupling (see Supporting Information). **Figure 2c** shows the calculated and experimental extinction spectrum of the NPs together with the calculated near-field intensity distribution $|E|^2 \propto |E_x(x,y)|^2/|E_0|^2$, when a NP is illuminated by $z$-polarized light (with field $E_0$) propagating in the $+x$ direction at $\lambda = 417$ nm (inset). The obtained spectral alignment with the polymer emission (indicated by green mark in **Figure 2b**) ensures an optimal condition for plasmon-exciton coupling and radiative rate enhancement in the composite films.[48]

Scanning transmission electron microscopy (STEM-HAADF) images of the Ag PNPs (**Figure 2e**), due to Z-contrast sensitivity, highlight well-defined cores of diameter of 28 ± 6 nm surrounded by 8 ± 3 nm thick $SiO_2$ shells. The synthesis route, first attained silver NPs by polyol reduction of silver salt[54], followed by silica shelling using a modified Stöber process[55], allowed shell thickness tuning without affecting Ag core integrity. These features are essential for achieving reproducible plasmonic resonance in hybrid composites. Energy-dispersive X-ray spectroscopy (STEM-EDX) elemental maps (**Figure 2e, f, Supplementary Fig.S3**) confirm that the spatial distributions of Ag, Si, and O are consistent with the observed core-shell architecture. Notice that the Ag PNAs retain identical plasmonic resonance peak even after integration into the polymer matrix, confirming that the silica shell effectively suppresses dielectric environment change and aggregation. In contrast, bare Ag NPs directly deposited in solid films exhibit a red-shifted and broadened plasmon band, ascribed to dielectric constant variation and interparticle coupling[56]; the comparative data are shown in **Supplementary Fig.S4**. The combined microscopic and spectroscopic data confirm the structural integrity and compositional purity of the chosen NPs, forming the foundation for subsequent optical and scintillation studies.

Once the individual components of our sensitized blends had been characterized, and before proceeding to the fabrication of the complete hybrid system, the demonstration of Purcell enhancement, and its radiometric validation, it was necessary to clarify how blending PFO with $HfO_2$ NPs alters its morphology/intermolecular aggregation and photophysics. It is well established that conjugated polymers processed in the solid state undergo substantial intermolecular aggregation driven by π–π stacking between adjacent backbones[57,58] . In PFO, this typically results in the formation of the poorly emissive crystalline β-phase, in contrast to the highly luminescent "glassy" α-phase, where chains adopt a more disordered conformation and remain spatially isolated. The increased chain planarity and aggregation associated with the β-phase promotes the formation of low-$\Phi_{PL}$ molecular dimers and enhances exciton migration toward nonradiative trap sites, thereby reducing both emission efficiency and spectral purity. In plastic electronics, mitigating such aggregation has been the subject of extensive research[49,59,60], yet controlling secondary interactions remains challenging, particularly in long-chain, non-ionic polymers such as PFO. These systems are not readily compatible with supramolecular encapsulation strategies relying on ionic interactions, such as wrapping by wide-bandgap polymers (e.g., polyethylene oxide, amylose)[61,62] or incorporation into interlocked structures like polyrotaxanes that use cyclodextrins as spacers[63].

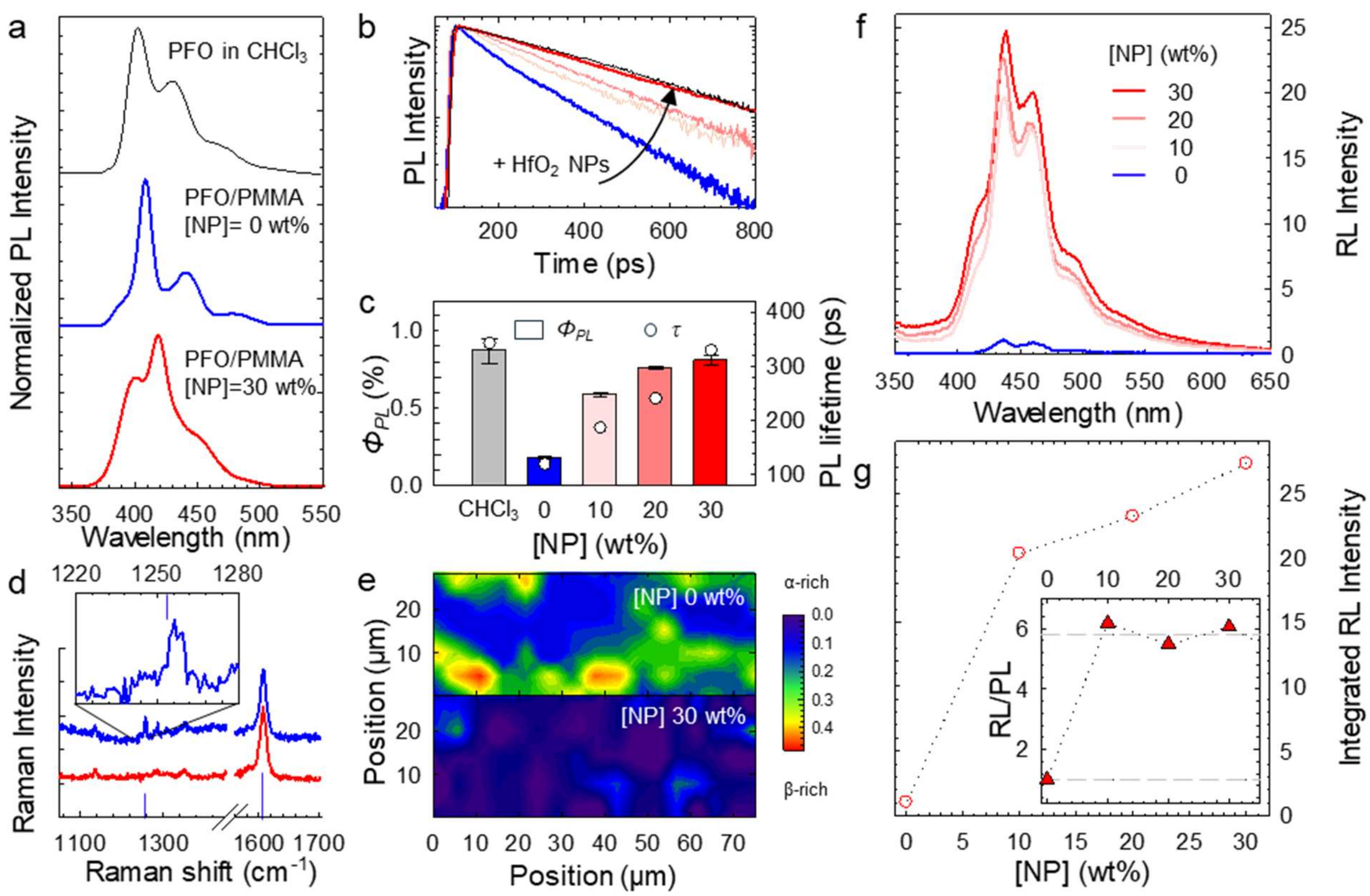

**Figure 3. Spectroscopic and radioluminescence characterization of PFO/PMMA–$HfO_2$ composites. a.** PL spectra of PFO in $CHCl_3$ solution (black), PFO/PMMA (blue), and PFO/PMMA films containing 30 wt% $HfO_2$ NP (red). **b.** Normalized PL decay curves of PFO/PMMA blends with increasing content of $HfO_2$ NPs (color code as in **c**). **c.** Corresponding PL quantum yields and decay times. **d.** Raman spectra highlighting changes in the 1255 $cm^{-1}$/1605 $cm^{-1}$ ratio, indicating reduced β-phase content upon incorporating 30 wt% $HfO_2$ NPs (lower panel) relative to bare PFO/PMMA blend (upper panel). The spectra are normalized and shifted vertically for clarity. Inset: enlarged view of 1255 $cm^{-1}$ peak, characteristic of β-phase PFO. **e.** Raman-mapping false color images comparing the spatial distribution of β-phase fraction in PFO/PMMA and PFO/PMMA–$HfO_2$ (30 wt%). **f.** RL spectra of the sample set (color code as in **c**) normalized to the RL intensity of the bare PFO/PMMA blend. **g.** Integrated RL intensity as a function of $HfO_2$ NP loading relative to the value for the bare PFO/PMMA blend. Inset: RL/PL intensity ratio, which isolates the net high-Z-sensitization enhancement of the scintillation intensity from the effect of increasing $\Phi_{PL}$. Above ≈10 wt% of NP, the RL/PL ratio saturates at ~ 6-fold increase, indicating that all emitting domains in the sample are activated under ionizing excitation.

Consistent with earlier studies, blending PFO with inert polymers proves largely ineffective except at extremely low PFO fractions[64], which would be unsuitable for scintillation due to insufficient emissive content - notice that even at extremely low concentrations, long-chained polymers can still bend and form intra-chain aggregates similar to β-sheets in macromolecular proteins[64]. In the absence of strong specific interactions (*e.g.* ionic forces or engineered steric blocking) polymer mixtures tend to phase-segregate, creating local domains enriched in one component and thus promoting aggregation. In agreement with these observations, casting neat PFO on glass leads to an ~80% reduction in $\Phi_{PL}$ with respect to the dilute solution (from 87% in solution to 18% in film), accompanied by a markedly faster nonradiative decay ($\tau_{film}$=120 ps) and a PL spectrum characteristic of the β-phase, with an emission peak at 437 nm (**Supplementary Figure S1**). The β-phase also manifests in absorption as a shoulder near 435 nm[61]. As shown by Tomoya *et al.*[65], even ~0.5% β-phase is sufficient to dominate the PL spectrum through the red-shifted 1-0 vibronic transition. Blending PFO into PMMA (1 wt% PFO) yields a minor improvement in optical performance, and further dilution would undesirably reduce emissive density. Remarkably, however, incorporation of $HfO_2$ NPs in the same blend produces a pronounced recovery (**Figure 3a** and **b**, color coded red). The PL spectrum becomes increasingly α-phase-like, and $\Phi_{PL}$ rises to ~80%, approaching that of isolated chains in solution - we notice that the lower intensity of the 0-0 vibronic peak with respect to the solution is likely due to partial reabsorption by the relatively thick film (1.8±0.1 μm) consistent with the observed PL excitation signal at ~400 nm (**Supplementary Fig.S5**). Time-resolved PL (**Figure 3b**) shows a systematic recovery of lifetime with increasing $HfO_2$ NP content, consistent with reduced nonradiative loss pathways and restoration of radiative dominated decay. A full comparison of $\Phi_{PL}$ and lifetimes for PFO/PMMA/$HfO_2$ NPs composite up to 30 wt% NP loading is presented in **Figure 3c**, which reveals that recovery of $\Phi_{PL}$ saturates at around 30 wt% of NP. This composition is therefore selected as the optimized NP loading for the subsequent Purcell-enhanced scintillator architecture. Raman spectroscopy provides complementary evidence for this morphological stabilization. Following Perevedentsev *et al.*[48], the ratio of the 1255 $cm^{-1}$ (in-plane C–H bending/C–C stretching) to the 1605 $cm^{-1}$ (aromatic ring stretching) modes serves as a quantitative marker of β-phase content. As shown in **Figure 3d**, the Raman spectrum of the PFO/PMMA blend exhibit a pronounced peak at 1255 $cm^{-1}$ peak consistent with substantial β-phase. Such signal is nearly absent upon addition of $HfO_2$ NPs. Raman mapping over a 50 μm × 100 μm area probing the β-phase through the intensity ratio between the 1255 $cm^{-1}$ and the 1605 $cm^{-1}$ lines (**Figure 3e**) confirms this effect: the ratio decreases markedly upon incorporation of $HfO_2$ NPs and reveals a spatially uniform α-phase distribution across the composite film. Additionally, we checked the intensity mapping of the characteristic 1605 $cm^{-1}$ line as an indicator of PFO concentration, and the result shows nearly uniform spatial distribution across both samples (**Supplementary Fig.S6**) indicating that phase separation between PMMA and PFO, if present, produces segregated regions smaller that the μ-Raman resolution (~0.5 μm). To confirm that the observed improvement in optical performance arises from the presence of the $HfO_2$ NPs, rather than solely from their organic capping ligands (BMEP and OAm), we also measured $\Phi_{PL}$ and PL lifetime of PFO/PMMA blends containing only the ligand mixture at the same concentration used in the $HfO_2$-loaded samples (see **Supplementary Fig.S7**). No measurable changes in the emission spectral shape, $\Phi_{PL}$, or decay dynamics were observed in these control samples. This result suggests that, in the absence of $HfO_2$ NPs, the molecular species phase-segregate in the solid state, thereby losing their beneficial role in suppressing interchain aggregation.

Importantly, the addition of $HfO_2$ NPs also has the crucial effect of enhancing the scintillation of the polymer blend through the photoelectrons generated when ionizing radiation interacts with the high-Z hafnium atoms. Under X-ray excitation, this high-Z sensitization acts in synergy with the improved luminescence efficiency, producing a dramatic increase of the RL output. As shown in **Figure 3f**, which reports the RL spectra of blends with and without $HfO_2$ NPs (measured for the same polymer mass and under identical excitation and collection conditions), the presence of the NPs leads to an approximately 30-fold increase in RL intensity. After accounting for the 4–5× difference in $\Phi_{PL}$ between the two extreme samples (namely without $HfO_2$ and with the highest loading in the set), the net high-Z sensitization effect amounts to roughly 6–7 fold with the addition of 30wt% of $HfO_2$ NPs. This trend is quantified in **Figure 3g**: the initial immediate 20-fold increase at 10 wt% addition comes from high-Z sensitization, whereas scintillation improvements after 10wt% are mainly attributed to increasing $\Phi_{PL}$. The RL/PL ratio (inset) saturates to ~ 6-fold enhancement above ≈10 wt%, indicating that in the employed conditions essentially all emissive domains are coupled to the high-Z sensitizer under ionizing excitation. We further note that the RL spectrum of the NP-free blend differs from its corresponding PL spectrum and more closely resembles the residual α-phase emission, suggesting that excitation of the aggregated β-phase is further disfavoured under ionizing excitation.

Having clarified the effect of high-Z sensitizers on both PL and RL, we now turn to the demonstration of Purcell enhancement under optical and ionizing excitation. For this stage, the PFO/PMMA/$HfO_2$ blend with [NP]=30wt% described in **Figure 3,** is used as the high-Z-sensitized control sample, while the same stock solution is supplemented with Ag-$SiO_2$ PNAs to produce the fully hybrid mixture exhibiting both high-Z sensitization and Purcell enhancement. The Ag-$SiO_2$ PNAs were incorporated into the PFO/PMMA blend concurrently with the $HfO_2$ NPs, yielding a homogeneous mixture identical in polymer and $HfO_2$ NP mass content to the control sample, apart from the addition of the plasmonic antennas. We stress that the amount of Ag-$SiO_2$ PNAs was only 0.05 wt%, orders of magnitude lower than the $HfO_2$

NPs, leading to negligible additional high-Z contribution to the RL enhancement with respect to the control sample. The influence of plasmonic coupling on exciton dynamics was first investigated under purely optical excitation, where the high-Z effect does not contribute.

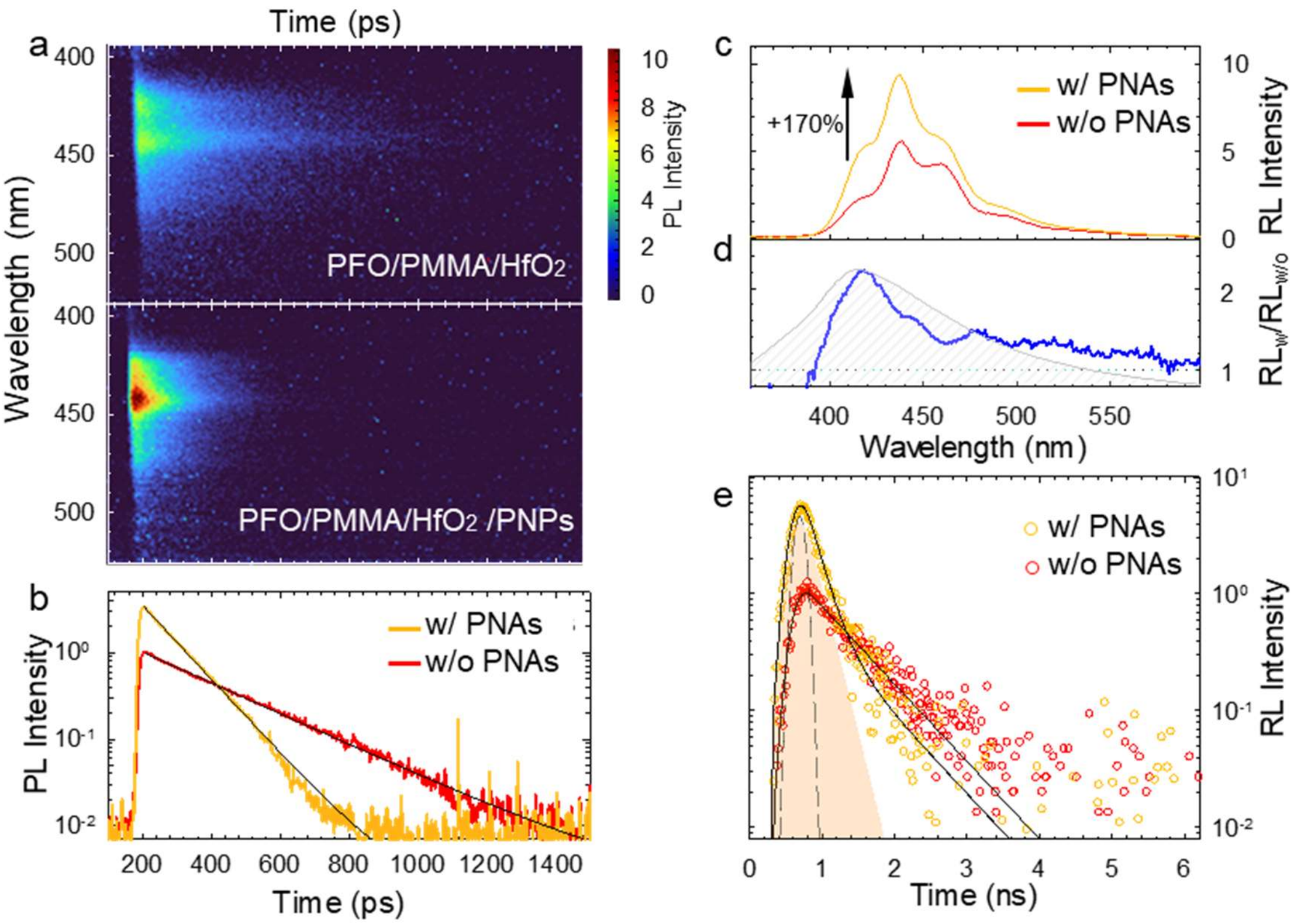

**Figure 4. Plasmonic enhancement of PL and RL performance in high-Z sensitized polymer scintillators. a.** Contour plots of the spectrally and time resolved PL intensity of PFO/PMMA/$HfO_2$ and Purcell-enhanced PFO/PMMA/$HfO_2$/PNAs nanocomposites measured under identical excitation and collection conditions. **b.** PL decay curves extracted from **a**. **c.** Steady-state RL spectra of the two nanocomposites under X-ray excitation. The RL enhancement factor $RL_w/RL_{wo}$ (blue) yields the wavelength-dependent amplification caused by plasmonic Purcell coupling, which matches the LSPR peak of Ag-$SiO_2$ PNP (shaded area). **e.** RL decay curves of the PFO/PMMA/$HfO_2$ (red circles)**,** PFO/PMMA/$HfO_2$/PNPs nanocomposites (orange circles) and the fitting functions (black lines); the shaded area indicates the plasmon enhanced contribution of the RL extracted from deconvolution with IRF (dotted line, see Supplementary Information for details), accounting for ~80% of the total emission in the Ag-$SiO_2$ PNA-doped sample.

**Figure 4a,b** show the contour plots of the time- and spectrally-resolved PL intensity for the control sample without PNAs (indicated as w/o PNAs) and for the mixture containing PNAs (indicated as w/ PNAs). In agreement with the data in **Figure 3**, the decay of the control sample is single-exponential, with a lifetime $\tau_{w/o}$=242 ps (**Figure 4c**). Considering the $\Phi_{PL}$=80%, this corresponds to a radiative lifetime of $\tau_{w/o}^{R}=\frac{\tau_{w/o}}{\Phi_{PL}}$=302 ps. The addition of PNAs substantially modifies the emission dynamics, leading to a markedly faster, single exponential, decay with a lifetime $\tau_w$ =100 ps. Under the assumption that the introduction of PNAs does not add new nonradiative decay channels (confirmed by the control experiments reported below), this corresponds to a radiative lifetime $\tau_w^R$ =106 ps and to an increased $\Phi_{PL}$=94%. Taken together, these data yield a Purcell factor, $F_P=\tau_{w/o}^R/\tau_w^R$=2.85, in agreement with previous reports on plasmon-enhanced polymer systems[21,35]. The small difference between the calculated $F_P$ and the experimentally observed increase of the zero delay PL intensity ($I_w/I_{w/o}$~3.2) indicates a minor additional contribution by optical-field enhancement by the PNAs. This interpretation is consistent with the observed ~1.5-fold increase of the integrated PL intensity upon addition of PNAs.

Crucially, under ionizing excitation the same behaviour persists. As is shown in **Figure 4c,** the blend with Ag-$SiO_2$ PNAs exhibits ~1.5-times higher spectrally integrated RL intensity with respect to the control sample when measured under identical excitation, detection, and collection geometries, which is consistent with the presence of field enhancement under ionizing radiation. The wavelength-dependent plasmonic amplification factor ($RL_w(\lambda)/RL_{w/o}(\lambda)$) is obtained by dividing the RL spectrum of the Purcell-enhanced film ($RL_w(\lambda)$) by that of the PNA-free reference ($RL_{w/o}(\lambda)$), revealing remarkable match between the Purcell enhanced RL and the LSPR spectrum of PNAs (**Figure 4d**). The scintillation yield was evaluated by direct comparison, under identical excitation and collection conditions, with a reference specimen of EJ276D plastic scintillator (100 µm thickness; *LY* = 8,600 ph/MeV, stopping power 4.5±0.4% of the incident X-ray beam). The comparative RL spectra are shown in **Supplementary Fig.S8**. The fraction of energy deposited in the nanocomposite films was measured following the methodology established for thin-film scintillators[66]. Briefly, the scintillation intensity of a bulk LYSO scintillator placed downstream of the nanocomposite samples along the incident radiation path was monitored. To minimize experimental error, measurements were carried out using dedicated hollow holders without any substrate that could contribute to beam attenuation. The measurements were performed by stacking a progressively increasing number of nanocomposite layers and extracting the deposited energy fraction through appropriate interpolation (**Supplementary Fig.S9**). The measurements yielded an attenuation value of the incident beam of 2.9±0.3% for the nanocomposite sample, consistent with the high loading of $HfO_2$ NPs (we notice that the presence or absence of PNAs has no measurable effect on the stopping power given their negligible concentration with respect to the $HfO_2$ NPs). After correcting for the respective fractions of absorbed X-rays, we obtained *LY* of 3,700±700 ph/MeV for the control sample without PNAs and as high as 6,100±1200 ph/MeV for the PNA-enhanced nanocomposite. The observed value rivals that of commercial plastic scintillators with significantly longer decay times (by over an order of magnitude, see **Figure 1a**) and exceeds ultrafast 'quenched' alternatives by more than tenfold. Crucially, a significant performance advantage persists even in the absence of PNAs, confirming the robust potential of sensitized conjugated polymer scintillators when interchain aggregation is suppressed.

The corresponding time-resolved RL decays display a comparable lifetime acceleration to the respective PL (**Figure 4e**). The control sample shows a single-exponential decay with lifetime of $\tau_{w/o}^{RL}$~440 ps, whereas PFO/PMMA/$HfO_2$/PNA nanocomposite displays a double-exponential RL decay with a $\tau_{w}^{RL}$~150 ps fast component accounting for ~80% of the RL signal and a slower one (~470 ps), associated minor portions of the film where plasmonic coupling is hindered. This component is thus expected to be eliminated by optimization of the film morphology. The average lifetime of the current test-bed sample, calculated as the harmonic average of the decay contributions weighted for the respective integrated weight is 215 ps, which is still less than one third of the lifetime of the fastest commercial scintillators available (~700 ps, EJ232Q, BC-422-5). As highlighted in **Figure 1a**, this approach enables one of the most favourable combinations of $LY$ and timing performance reported for plastic scintillators to date. The ratio between the measured RL lifetime components yields an effective $F_{P,RL}$~2.9, in line with the optical results. We clarify that the difference between the observed PL and RL lifetimes likely arises from the different experimental setup: in RL mode, excitation is provided by ~100 ps X-ray pulses and detection is performed with a photomultiplier tube with a ~50 ps instrument response, resulting in a ~120 ps IRF, as opposed to a streak camera measurement pumped by ~150 fs laser pulses. The PL decay collected with the same setup as the RL is reported in **Supplementary Fig.S10** for clarity. For completeness, we conducted control experiments in which the plasmonic field of the PNAs was intentionally decoupled from the PFO moieties. This was achieved either by increasing the silica-shell thickness to spatially separate the components (**Supplementary Fig.S11)** or by selecting a polymer whose emission is off-resonance with the LSPR of the PNAs (**Supplementary Fig.S12** and **S13**). In both cases, no measurable enhancement or quenching in either PL or RL was observed. These results confirm that the enhancements and radiative-rate acceleration reported in Figure 4 arise from genuine plasmonic coupling and confirm that the addition of PNAs does not introduce additional nonradiative quenching channels.

In summary, we demonstrated the first example of fully solution-processable hybrid scintillator that combines three synergistic mechanisms: intrinsic ultrafast luminescence of conjugated polymers, high-Z sensitization, and Purcell-enhanced radiative recombination, within a single nanocomposite platform. By integrating transparent high-Z sensitizers and plasmonic nanoantennas into a conjugated polymer matrix, we simultaneously suppressed aggregation-induced quenching of the conjugated emitters, enhanced energy deposition under ionizing radiation, and accelerated radiative decay via plasmon-exciton coupling. Importantly, the Purcell effect was shown to operate equivalently under both optical and X-ray excitation, directly linking plasmon-enhanced exciton recombination to scintillation dynamics. Although this proof-of-concept demonstration of the first plasmonic enhanced scintillator with largely sub-ns lifetime and competitive efficiency is not intended to represent a final device architecture, and further optimization is clearly possible (e.g., fine-tuning nanoparticle size and concentration, improving film morphology, or extending the approach to thicker composites for integrating into metascintillator schemes), these results establish a promising pathway toward high-performance, ultrafast organic scintillators. The demonstrated strategy is fully scalable, compatible with solution processing, and directly applicable to emerging detector architectures requiring efficiency, sub-nanosecond timing, scalability, and flexible form factors. More broadly, this work provides quantitative design guidelines for integrating plasmonic and high-Z nanomaterials with conjugated polymers, opening new opportunities for next-generation hybrid scintillators for high-energy physics, medical imaging, and advanced time of flight radiation-sensing technologies.

**Acknowledgments**

This work was funded by Horizon Europe EIC Pathfinder program through project 101098649 – UNICORN, by the European Union -Next Generation EU, Mission 4 Component 1 CUP H53D23004670006, and through the Italian Ministry of University and Research under PNRR-M4C2-I1.3 Project PE_00000019 "HEAL ITALIA". This research is funded and supervised by the Italian Space Agency (Agenzia Spaziale Italiana, ASI) in the framework of the Research Day "Giornate della Ricerca Spaziale" initiative through the contract ASI N. 2023-4-U.0t.

**Competing interests**

The authors declare no competing interests.

***Supplementary Information***

**High-Yield, 150-ps Polymer Scintillators via Synergistic Purcell and High-Z Enhancement**

Xiaohe Zhou[1], Hiba H. Karakkal[1], Chenger Wang[1], Andrea Fratelli[1,3], Francesco Bruni[1,2], Matteo L. Zaffalon[1,2], Leonardo Poletti[4], Saptarshi Chakraborty[1], Emanuele Mazzola[5], Roberto Lorenzi[1], Francesco Carulli[1], Francesca Rossi[4], Francesco Meinardi[1], Luca Gironi[2,5] and Sergio Brovelli*[1,2]

[1] *Dipartimento di Scienza dei Materiali, Università degli Studi di Milano-Bicocca, Via R. Cozzi 55, 20125, Milano, Italy*

[2] *INFN - Sezione di Milano - Bicocca, Milano 20125 – Italy*

[3] *Nanochemistry, Istituto Italiano di Tecnologia, Via Morego 30, 16163, Genova, Italy*

[4] *IMEM-CNR, Parco Area delle Scienze 37/A - 43124 Parma, Italy*

[5] *Dipartimento di Fisica, Università degli Studi di Milano-Bicocca, Piazza della Scienza, 20125 Milan, Italy*

Email: sergio.brovelli@unimib.it

* Corresponding author

## Materials and methods

### Synthesis of Silica coated silver nanoparticles

<u>*Chemicals:*</u> Silver nitrate ($AgNO_3$,≥99.0%, Sigma-Aldrich), polyvinyl pyrrolidone (PVP, average molecular weight ~55000), ethylene glycol (EG, anhydrous, 99.8%, Sigma-Aldrich), tetraethyl orthosilicate (TEOS, ≥99.0%, Fluka) , aqueous ammonium hydroxide solution ($NH_4OH$, ~30%, Sigma-Aldrich), octadecyltrichlorosilane (OTS, 95%, Acros Organics), ethanol (absolute, Sigma-Aldrich) acetone ((≥99.8%, Fischer chemicals), toluene (anhydrous, 99.8%, Sigma-Aldrich). All reagents were used as purchased.

<u>*Synthesis of PVP-capped silver nanoparticles (AgNP):*</u> Silver nanoparticles of average diameter 32nm were synthesized via polyol reduction of silver nitrate, using ethylene glycol as the high boiling solvent which also served as the reducing agent for silver salt along with PVP as the capping/stabilizing polymer to control particle size, and to prevent the aggregation. Briefly, 1.5g of PVP was dissolved in 6 mL ethylene glycol in a 50 mL round-bottom flask by sonication and stirring, followed by the addition of 100 mg of $AgNO_3$. The resultant mixture was heated to 125°C for 1 hour. The resulting PVP-stabilized silver nanoparticles were purified by washing with excess acetone as antisolvent to remove unreacted precursors and excess PVP and subsequently collected by centrifugation.

<u>*Silica coating of silver nanoparticles (Ag-SiO₂):*</u> Silica-coated silver nanoparticles were prepared using a modified Stober process by stepwise addition of TEOS. PVP-stabilized silver nanoparticles were first redispersed in 140 mL of ethanol, followed by the addition of 4 mL ammonium hydroxide solution under stirring. To this mixture, 750 µL of TEOS was introduced and the solution was stirred for 30 minutes, after which a second aliquot of 750 µL TEOS was added. Following the second TEOS addition, half of the reaction mixture was extracted out, centrifuged at 7830 rpm for 10 minutes, and the resulting precipitate was redispersed in water. The particles were subsequently washed by repeated centrifugation in water and ethanol. The remaining portion of the reaction mixture was allowed to react for an additional 3 hours to promote the growth of a thicker silica shell, and was collected and purified in the same manner as the first batch. The resultant particles in each case were redispersed in 5 mL ethanol.

*Synthesis of $HfO_2$ NPs:* $HfO_2$ NPs were synthesized by the thermal decomposition of $Hf(CF_3COO)_4$ at high temperature following a modified version of the procedure reported by Pei *et al.*[1] and our previous work[2]. The precursor $Hf(CF_3COO)_4$ was obtained by dissolving 9.6 g of $HfCl_4$ in 50 ml of trifluoroacetic acid at 40 °C overnight, which was followed by drying the white reaction mixture in a rotary evaporator to obtain the desired white powdered $Hf(CF_3COO)_4$. 3.15 g of this precursor in 50 ml of oleylamine (OAm) was degassed at 110 °C under vacuum for one hour, following which the reaction mixture was taken to 330 °C under an inert atmosphere. After approximately 30 minutes, the solution turned colorless, indicating particle nucleation. The particles were allowed to grow for 15 minutes, after which the reaction flask was removed from the heating mantle and allowed to cool down to room temperature naturally. The NPs were initially precipitated by flocculating with 120 ml of acetone and centrifuging at 7830 rpm for 3 minutes, and later by re-dissolving in 40 ml of toluene and flocculating with 160 ml of ethanol three times. The obtained white NCs were dried under vacuum and dissolved in chloroform. Typically, 1.3 g of $HfO_2$ was obtained per synthesis. For polymer nanocomposite preparation, the native OAm shell was partially exchanged with bis(2-(methacryloyloxy)ethyl) phosphate (BMEP) to introduce polymerisable methacrylate groups and improve nanoparticle–matrix bonding. Ligand exchange was performed by stirring the OAm-capped $HfO_2$ NPs with a controlled amount of BMEP in chloroform overnight, followed by repeated precipitation and redispersion. As discussed by Pei *et al.*, the degree of surface modification can be tuned via the BMEP/$HfO_2$ feed ratio; in this work we used 50% BMEP exchange, corresponding to an approximately 1:1 molar ratio of OAm and BMEP ligands on the NP surface, which was previously identified as an optimal compromise between solubility and surface monomer density. The final organic composition in our samples our NPs contain 84 wt% $HfO_2$, 9 wt% BMEP and 7 wt% OAm.

*Film Fabrication*. PFO and PMMA were dissolved in chloroform at total concentration 1 mg $mL^{-1}$ , maintaining the reported PFO:PMMA ratio. $HfO_2$.NP (10–50 wt %) and Ag PNA [ Ag 0.05wt % ] were dispersed in ambient air by ultrasonication. Films were deposited by dropcasting on glass/PTFE substrate.

*Morphological and elementary characterization.* The powder X-ray diffraction (XRD) patterns of samples were performed by a Bruker D8 Advance X-ray Diffractometer at 40 kV and 30 mA using Cu Kα radiation (λ = 1.5406 Å). TEM/STEM imaging and EDX spectroscopy were performed in a JEOL JEM-2200FS microscope, operated at 200 kV, equipped with a high-angle annular dark field detector for Z-contrast imaging and an Oxford Xplore detector for compositional analysis. The particles were deposited by drop-casting on Cu grids with ultrathin carbon support film.

*Thermogravimetric analysis.* Evaluation of $HfO_2$/organic ligand ratio: The quantification of the fraction of inorganic material in BMEP-exchanged $HfO_2$ NPs was performed using the thermo-gravimetric analysis (TGA). The sample was prepared depositing ca 5 mg of BMEP-exchanged $HfO_2$ NPs from a $CHCl_3$ solution in a silica crucible. The measurement was performed at constant air flux (50 mL $min^{-1}$) in the range of temperature 30-1000°C with a constant 10°C/min heating ramp. The fraction of inorganic material was determined from the ratio between sample weight above 700°C compared with the weight at 120°C (in order to avoid weight losses related to residual solvent evaporation). TGA analyses were carried out using a TGA/DCS1 STARe SYSTEM (Mettler Toledo, Columbus, OH, USA)

*Optical spectroscopy.* Optical absorption measurements were measured in octane with an Agilent Cary 60 UV–Vis spectrophotometer. PL measurements were performed by exciting the samples with a 405 nm pulsed diode laser (Edinburgh Inst. EPL 405, 70 ps pulse width) and collected with a TM-C10083CA Hamamatsu Mini-Spectrometer. PL quantum yield for every sample was obtained by comparison with a standard with the same absorbance at the excitation energy. With the same excitation source and spectrometer detection setup for steady-state PL measurement. For low fluence ultrafast time-resolved PL (TRPL) measurements, the samples excited by frequency-doubled Ti:sapphire laser ($E_{exc}$ = 3.26 eV, pulse duration ~150 fs, repetition rate ~ 76 MHz), the emitted light is collected with Hamamatsu streak camera (time resolution < 10 ps).

*Raman spectroscopy measurement* Raman maps were collected using a Labram Dilor spectrometer coupled to an Olympus BX40 optical microscope with a 10x objective lens resulting in a spectral and lateral resolution of about 1 $cm^{-1}$ and 1 μm, respectively. Spectra were acquired in backscattering configuration using a 633 nm He-Ne laser as the excitation source. Raman maps were obtained on a square matrix of points on a 100 μm x 100 μm area, with a step of 1 μm. Maps were reconstructed through the analysis of the ratio between the signal at 1250 $cm^{-1}$ and that at 1605 $cm^{-1}$.

*RL measurements* Unfiltered X-rays were produced using a Philips PW2274 X-ray tube with a tungsten target, equipped with a beryllium window and operated at 20 kV to produce a continuous X-ray spectrum through bremsstrahlung. The scintillation light was detected using a liquid-nitrogen-cooled, back-illuminated, UV-enhanced CCD detector (Jobin Yvon Symphony II), coupled to a monochromator (Jobin Yvon Triax 180) with a 100 lines/mm grating.

*LY measurements* Light yield values were determined by comparing the integrated RL intensity under 20 kV X-ray excitation ($\langle E \rangle$ ~ 9 keV) with identical experimental conditions for a 1 wt% octane solution of $CsPbBr_3$ NCs and NC-MSNs placed in a 5 mm long crucible and a commercial EJ-276D plastic scintillator (LY = 8600 photons/MeV) of the same size and geometry used as a reference. In both cases, the sample size was chosen to ensure complete attenuation of the excitation beam.

*Time resolved scintillation measurements* The time-resolved RL was measured using a pulsed X-ray source consisting of a 405 nm ~70-ps pulsed laser hitting the photocathode of an X-ray tube (N5084, Hamamatsu) set at 40 kV. The emitted scintillation light was collected using an Horiba MicroHR (grating 150 lines/mm) coupled to a PMA hybrid photomultiplier from PicoQuant operated in single photon counting mode in combination with a PicoHarp 300 time correlator. The RL decay curves were analyzed using a least-squares fitting approach with the following formula, which accounts for the convolution with the instrument response function (IRF):

$$F(t) = IRF(t) \otimes \left( H(t - t_0) \cdot \left[ \sum_{i=1}^{2} a_i \cdot e^{-t/\tau_i} \right] \right) + C$$

where $t_0$ corresponds to the start of the emission process, *C* is the electronic background noise floor, and *H* is the Heaviside function. The experimental IRF was well described by a Gaussian profile (FWHM = 120 ps), and the weight of each component ($w_i$) was calculated as the integral of each convoluted function over the entire time window. The average lifetime was calculated using the re-normalized ratio of all components $\tau_i$ according to:

$$\tau_{eff} = \left( \frac{\tau_1}{w_{1n}} + \frac{\tau_2}{w_{2n}} \right)^{-1}, \qquad w_{in} = \frac{w_i}{w_1 + w_2}$$

_Electrodynamic model of plasmonic resonance:_ To provide an indicative estimate of the localized surface plasmon response of Ag-$SiO_2$ core–shell nanoparticles, we performed a simplified electromagnetic simulation in the dipole/quasi-static limit. The sampled wavelengths ($\lambda$) are considered in vacuum and converted to photon energy as $E(\lambda) = hc/\lambda$, where $h$ is Planck's constant and *c* is the speed of light in vacuum. The surrounding medium was described by a real refractive index $n_{\text{med}}$ (assumed dispersionless and fixed at $= 1.5021$), corresponding to a relative permittivity $\varepsilon_m = n_{\text{med}}^2$, and wavevector $k(\lambda) = 2\pi n_{\text{med}}/\lambda$. The $SiO_2$ shell was modelled as a lossless (real) dielectric with refractive index $n_{\text{shell}} = 1.45$ and relative permittivity $\varepsilon_2 = n_{\text{shell}}^2$. The Ag core complex dielectric function $\varepsilon_1(\lambda)$ was modeled with a Drude form

$$\varepsilon_1(E) = \varepsilon_\infty - \frac{\omega_p^2}{E^2 + i\gamma E}$$

where $\varepsilon_\infty$ is the high-frequency permittivity, $\omega_p$ is the plasma energy, $\gamma$ is the damping energy, respectively set at 5.0, 9.0 eV, and 0.02 eV, consistently with commonly used Drude parameterizations for silver and with experimental Drude analyses[3,4].

The nanoparticle geometry was defined by the Ag core radius *a* and the outer radius *b*=*a*+*t*, where *t* is the $SiO_2$ shell thickness. The core-to-particle volume ratio was $f = (a/b)^3$. Within the quasi-static approximation for concentric spheres, the electrostatic core–shell polarizability $\alpha_0(\lambda)$ was computed as:

$$\alpha_0(\lambda) = 4\pi b^3 \, \frac{(\varepsilon_2 - \varepsilon_m)(\varepsilon_1 + 2\varepsilon_2) + f(\varepsilon_1 - \varepsilon_2)(\varepsilon_m + 2\varepsilon_2)}{(\varepsilon_2 + 2\varepsilon_m)(\varepsilon_1 + 2\varepsilon_2) + 2f(\varepsilon_2 - \varepsilon_m)(\varepsilon_1 - \varepsilon_2)}$$

with $\varepsilon_1 = \varepsilon_1(\lambda)$, $\varepsilon_2$and $\varepsilon_m$as defined above. To enforce physical consistency between extinction and scattering, we included radiation damping (radiative reaction) via the corrected, frequency-dependent polarizability

$$\alpha(\lambda) = \frac{\alpha_0(\lambda)}{1 - i\dfrac{k(\lambda)^3}{6\pi}\alpha_0(\lambda)}$$

Optical cross sections were then evaluated in the dipolar limit as

$$C_{\mathrm{ext}}(\lambda) = k(\lambda)\,\mathrm{Im}[\alpha(\lambda)],$$

$$C_{\mathrm{sca}}(\lambda) = \frac{k(\lambda)^4}{6\pi}|\alpha(\lambda)|^2,$$

$$C_{\mathrm{abs}}(\lambda) = C_{\mathrm{ext}}(\lambda) - C_{\mathrm{sca}}(\lambda)$$

The resonance wavelength $\lambda_{\mathrm{res}}$ was taken as the maximum of $C_{\mathrm{ext}}(\lambda)$. Near-field enhancement was estimated at $\lambda_{\mathrm{res}}$ by representing the particle as an equivalent point dipole with moment $p = \alpha(\lambda_{\mathrm{res}})E_0$, excited by a uniform incident field of amplitude $E_0$ (normalized at 1 for clarity) along the x-axis. The scattered field was computed using the quasi-static dipole expression in real space, and the enhancement reported as $|E(\boldsymbol{r})|^2/|E_0|^2$, masking points inside the outer radius *b*.

*Determination of Ag mass fraction from LSPR absorption.* The weight percentage of silver in the Ag–$SiO_2$ PNP containing film was estimated from the UV–vis extinction at the localized surface plasmon resonance (LSPR) of the Ag cores. As is reported in the work by Kobayashi *et al.* [5], we use the extinction spectra of citrate-stabilized Ag and Ag-$SiO_2$ particles with a known Ag concentration (0.018 mM) and core size (~10 nm) as a calibration for the absorbance per unit Ag concentration at the LSPR peak. For our samples, the absorbance at the LSPR maximum $A_{sample}(\lambda_{LSPR})$ was measured in dilute dispersion in chloroform in 1 cm*1 cm quartz cuvette. By Beer–Lambert law, the Ag concentration was obtained as

$$c_{Ag} = c_{ref}\frac{A_{sample}(\lambda_{LSPR})}{A_{ref}(\lambda_{LSPR})}$$

where $c_{ref}$ and $A_{ref}$ are the Ag concentration and LSPR absorbance of the calibration data. The corresponding Ag mass loading was then calculated from $m_{Ag}=c_{Ag}VM_{Ag}$ and expressed as a weight fraction relative to the total solid content in the nanocomposite (polymer + $HfO_2$ + Ag–$SiO_2$). This procedure yields an estimated 0.05 wt% Ag in the final film.

**Supplementary Table 1:** Light yield and lifetime values for representative commercial (blue background) and research-grade (pink background) plastic scintillators.

| Scintillator Material | Decay Time (ps) | Light Yield (ph/MeV) | Key Feature | Reference |
|---|---|---|---|---|
| Anthracene | 30000 | 15625 | Standard reference | |
| Hybrid PFO/HfO2/PNA | 215 | 6100 | Conj. Polymer + High-Z + Purcell-effect | This work |
| Eljen EJ-200 | 2100 | 10000 | General purpose | https://eljentechnology.com/products/plastic-scintillators/ej-200-ej-204-ej-208-ej-212 |
| Luxium BC-408 | 2100 | 10000 | General purpose | https://luxiumsolutions.com/radiation-detection-scintillators/plastic-scintillators/bc400-bc404-bc408-bc412-bc416 |
| Eljen EJ-204 | 1800 | 10625 | General purpose | https://eljentechnology.com/products/plastic-scintillators/ej-200-ej-204-ej-208-ej-212 |
| Luxium BC-404 | 1800 | 10625 | General purpose | https://luxiumsolutions.com/radiation-detection-scintillators/plastic-scintillators/bc400-bc404-bc408-bc412-bc416 |
| Luxium BC-422 (Q) | 1600 (700) | 8400 (2900-460) | Fast timing | https://luxiumsolutions.com/radiation-detection-scintillators/plastic-scintillators/fast-timing-bc-418-bc-420-bc-422-bc-422q |
| Eljen EJ-232(Q) | 1600 (700) | 8400 (2900-460) | Fast timing (quenched) | https://eljentechnology.com/products/plastic-scintillators/ej-232-ej-232q |
| Eljen EJ-276D | 13000 | 8750 | Pulse shape discrimination | https://eljentechnology.com/products/plastic-scintillators/ej-276 |
| PVT/MF/SF/PBD/POPOP | 2300 | 11600 | Research plastic scintillators | [3] |
| PVT/ PPO/POPOP | several ns | ~commercial plastic scintillator | Research plastic scintillators | [4] |
| PS/POPOP/$HfO_2$ | 3690 | 10000 | High-Z sensitized | [14] |
| $Bi_2O_3$/PVT | 1300 | 5200 | High-Z sensitized | [15] |
| TPE-4Br@PVT | 1660 (fast) | 14443 | Hot-exciton plastic | [22] |

| | 5260 (slow) | | | |
|---|---|---|---|---|
| Conjugated-polymer scintillators | 1000 ~ 3000 | 2000~10 000 | Research plastic scintillator s | [23] |
| ZnO:Ga@PS 1 mm | 1200 | 500 | High-Z sensitized | [24] |
| Zr-DPA MOF | 7700 | 1160 | High-Z sensitized | [25] |

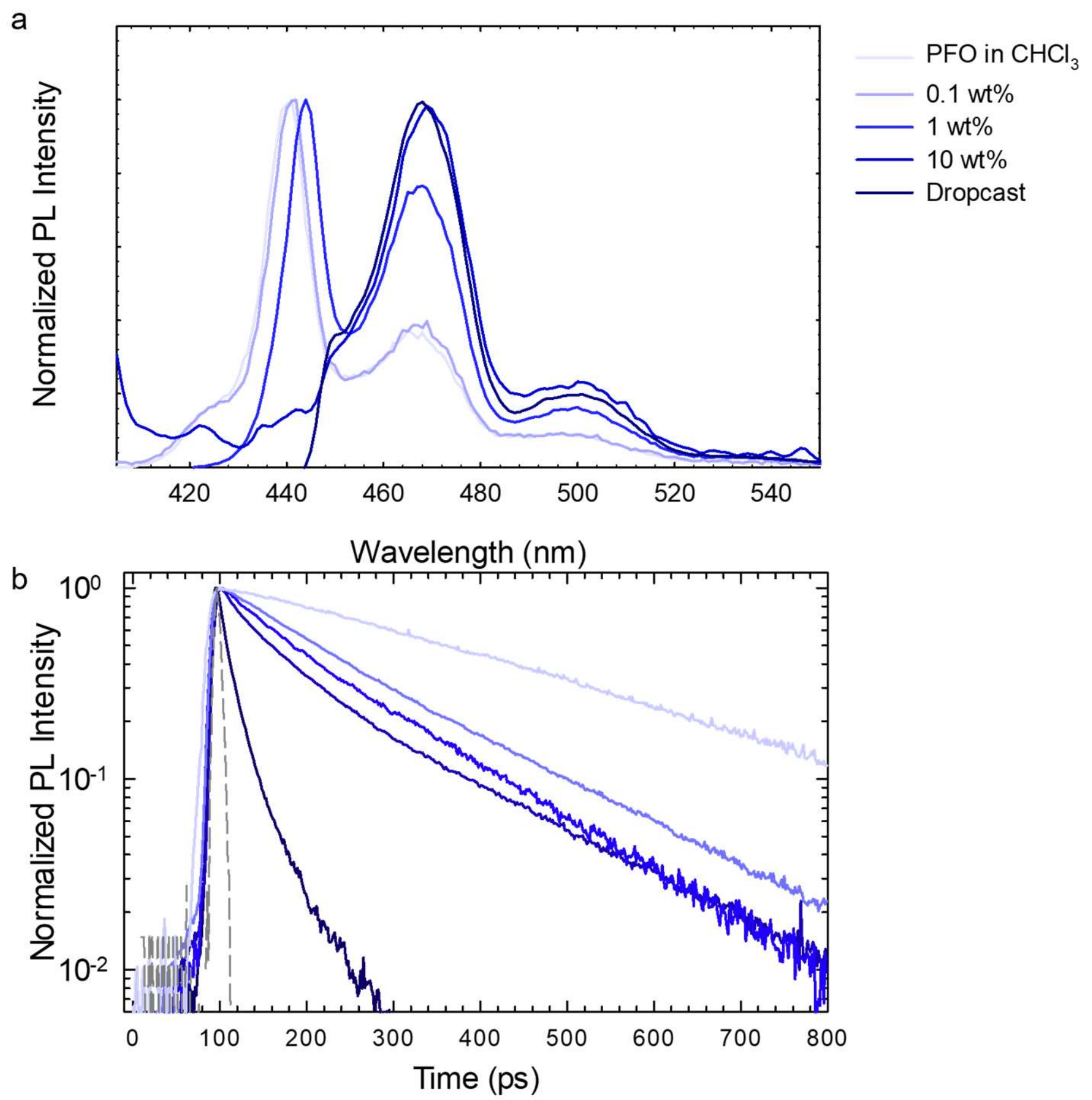

**Figure S1. a.** Representative PL spectrum of PFO/PMMA composite with different PFO loading. PFO solution (light blue) and PFO/PMMA composites with PFO weight percentages ranging from 0.01 wt% (blue) to 100 wt% dropcast PFO films. (dark blue).**b.** Streak camera–measured lifetimes of corresponding sample set.

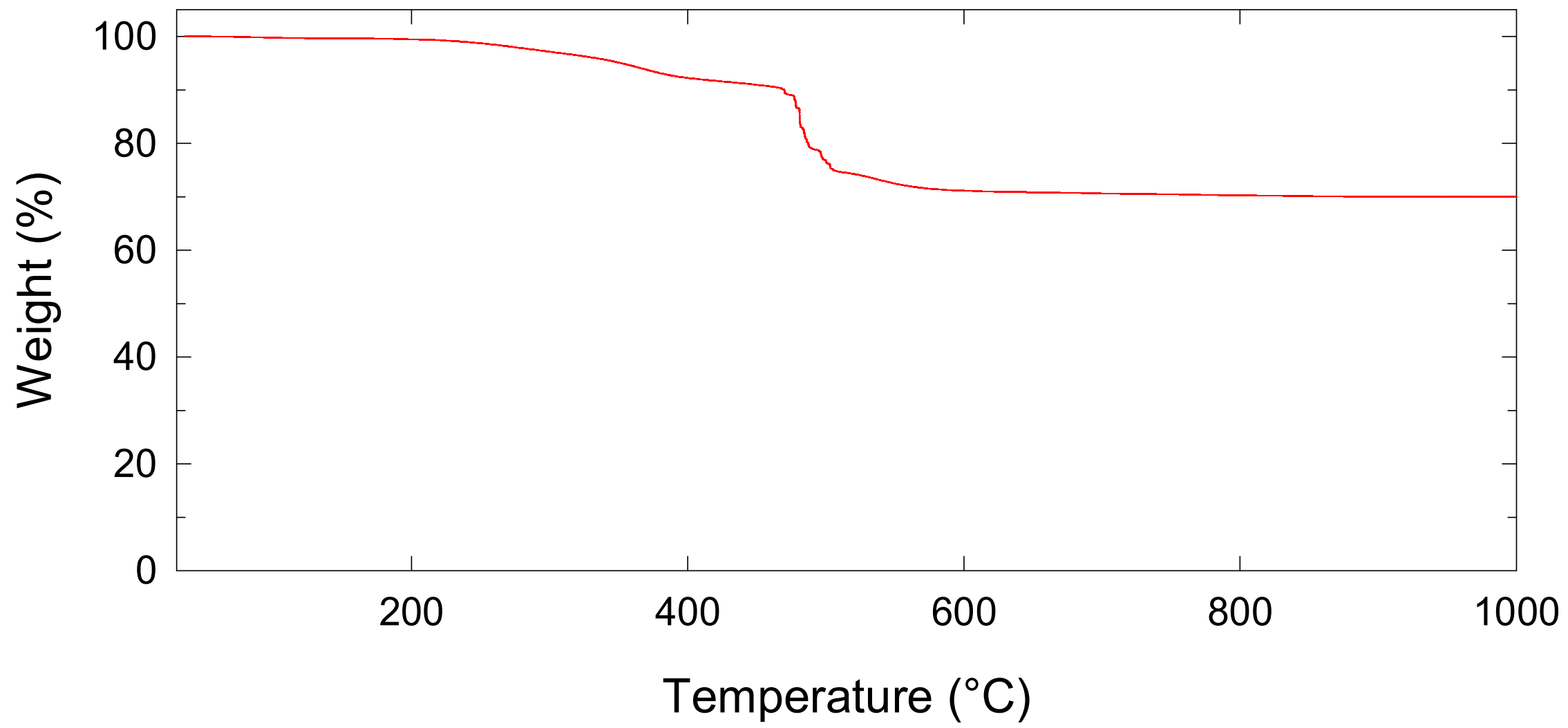

**Figure S2.** Thermogravimetric Analysis (TGA) of $HfO_2$ NPs collected in the range 30-1000°C with a heating rate of 10°C/min. The analysis highlights that the inorganic fraction of the NPs (extracted from the weight of material at high temperature) accounts for 70% of the total weight.

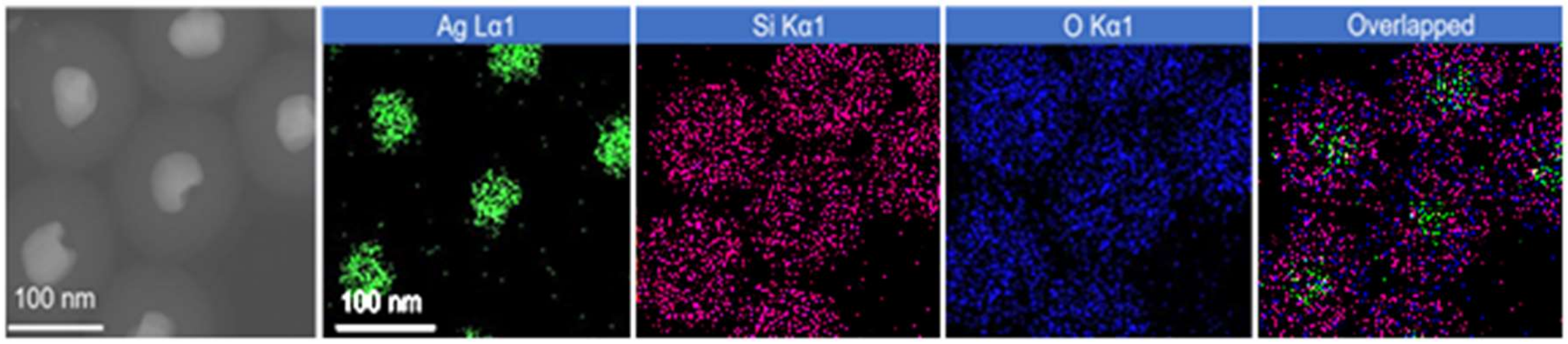

**Figure S3** STEM-HAADF image and elemental mapping analysis of synthesized Ag-$SiO_2$ nanoparticles with 30nm shell thickness.

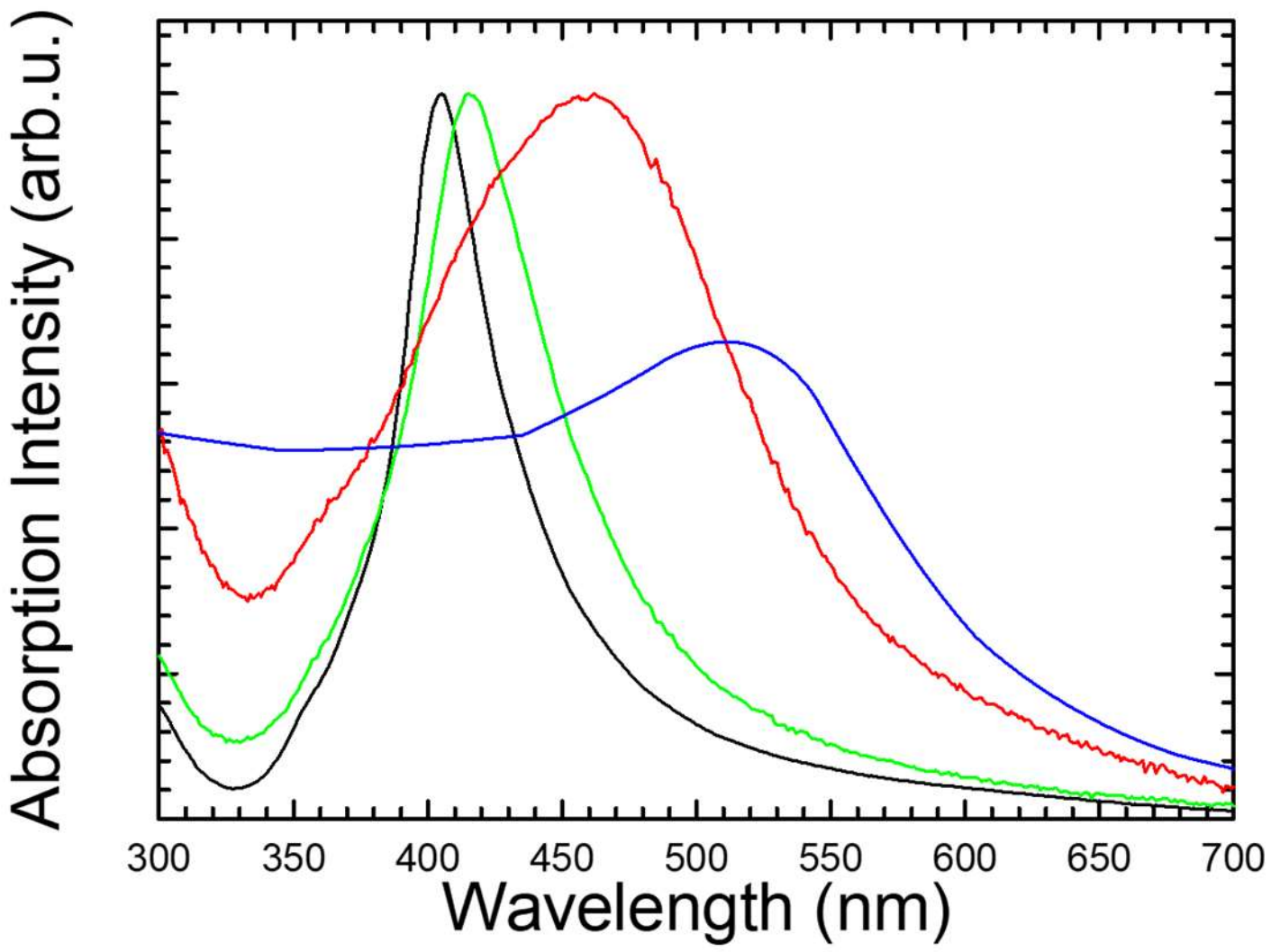

**Figure S4** Absorption profile of (black) bare Ag NP in water, (green) Ag-$SiO_2$ PNA coated with a 10 nm silica shell (the plasmonic antenna used in the main text), red) Ag-$SiO_2$ NP coated with a 30 nm silica shell and (blue) bare Ag NP casted on glass substrate. The progressive red-shift and broadening of the localized surface plasmon resonance (LSPR) peak with increasing shell thickness confirms successful silica encapsulation while preserving the Ag core and its dielectric environment.

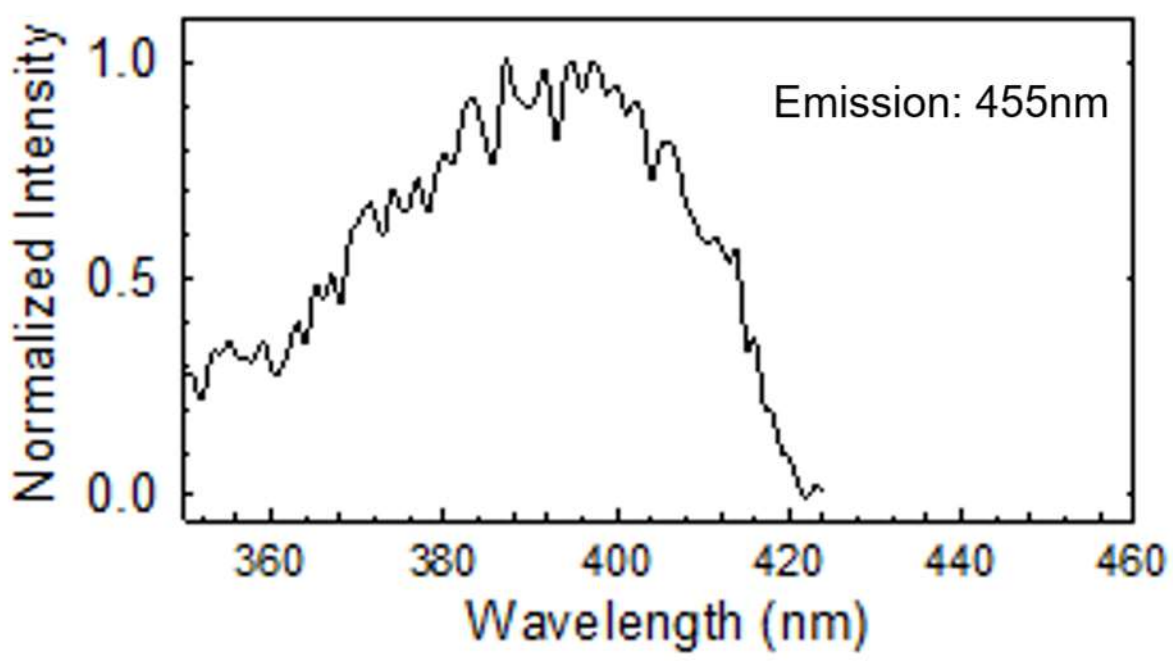

**Figure S5** PL excitation spectrum of PFO/PMMA/$HfO_2$ sample reported in the main text.

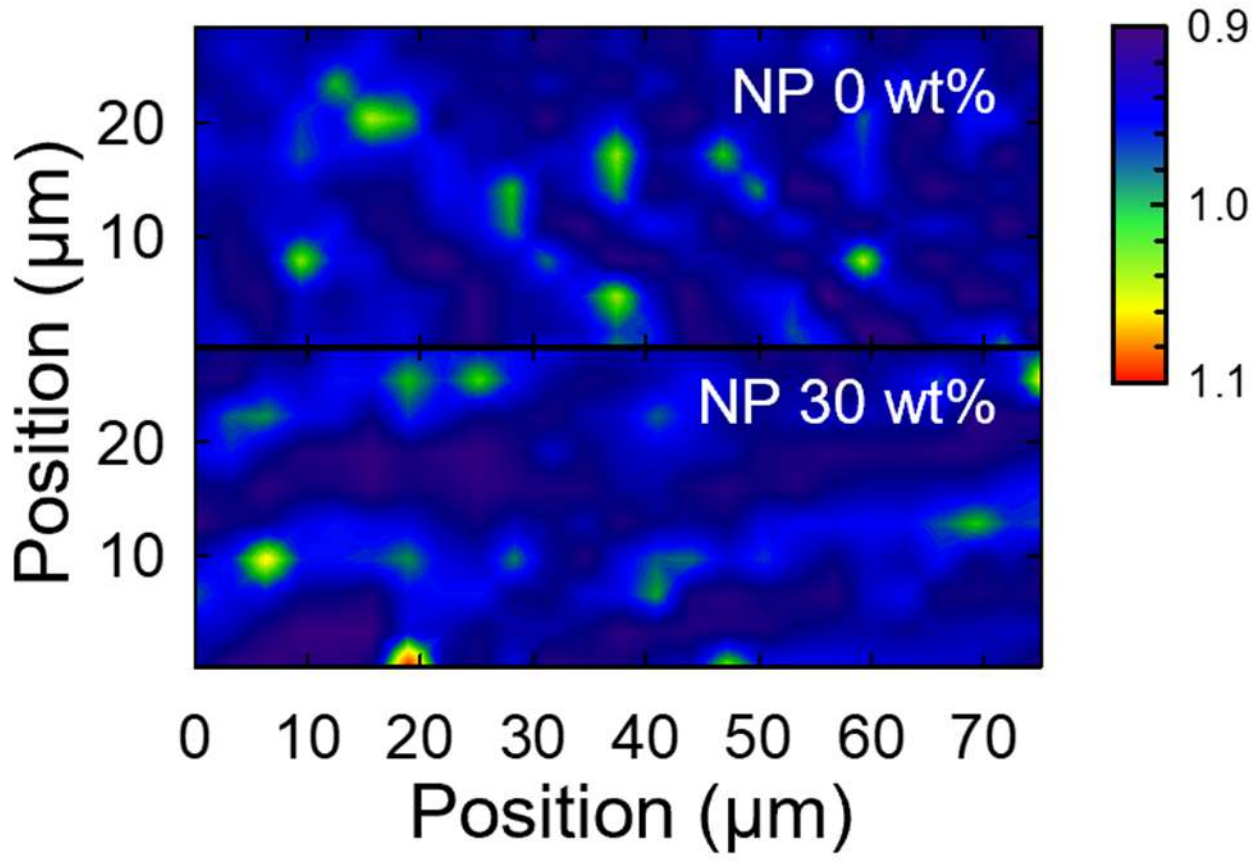

**Figure S6.** Raman mapping of PFO distribution in the polymer composite. Two-dimensional Raman map of the normalized baseline-corrected intensity of the PFO aromatic C=C stretching band at 1605 $cm^{-1}$, of the same region as shown in the main text. The nearly uniform intensity across the scanned area indicates a homogeneous distribution of PFO within the film under the applied excitation and collection conditions.

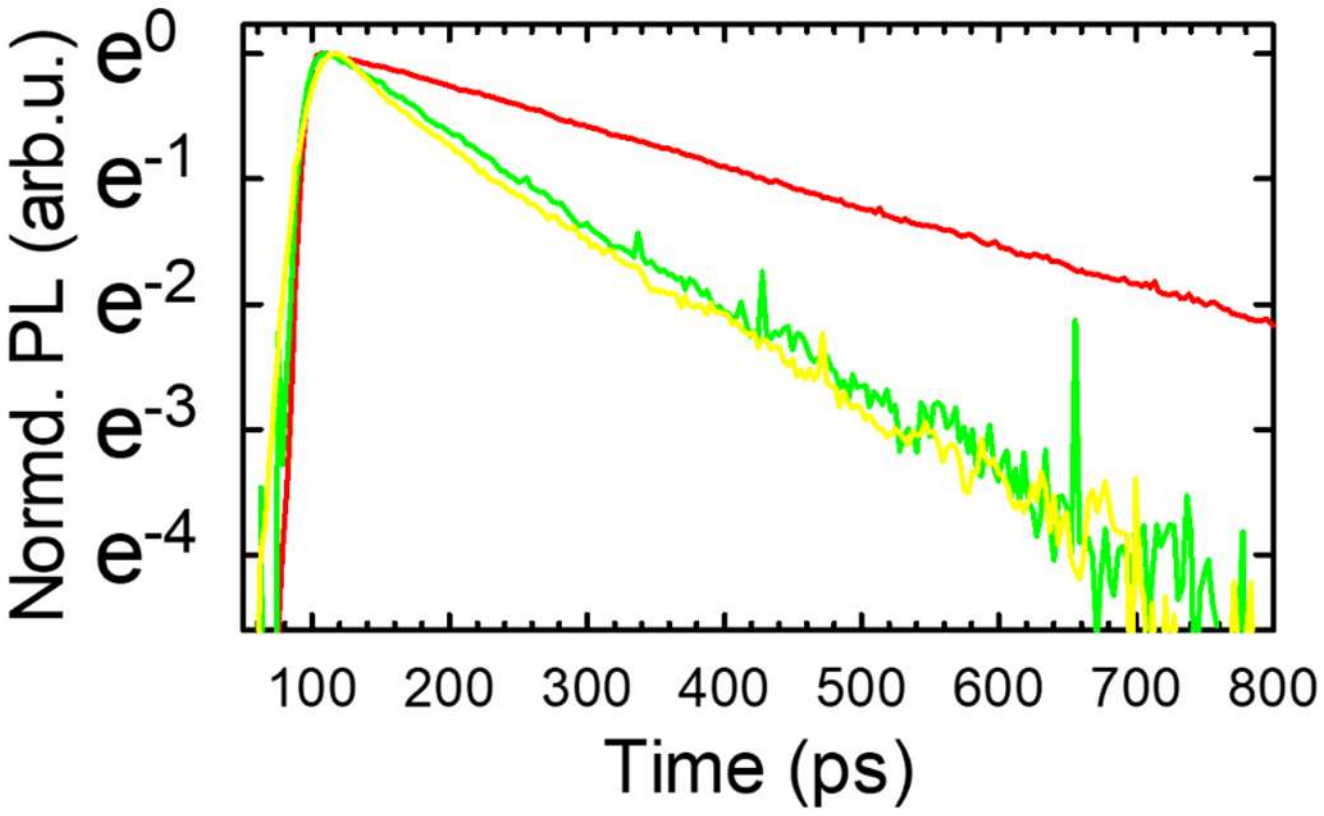

**Figure S7** TRPL dynamics disentangling ligand and $HfO_2$ contributions. Time-resolved photoluminescence decays extracted from streak-camera measurements for PFO/PMMA films containing (green) a mixture of BMEP and OAm ligands, (yellow) OAm only, and (red) 30 wt% $HfO_2$ NP. The ligand-only samples were prepared with the same weight percentage as BMEP/OAm content as in $HfO_2$ NP.

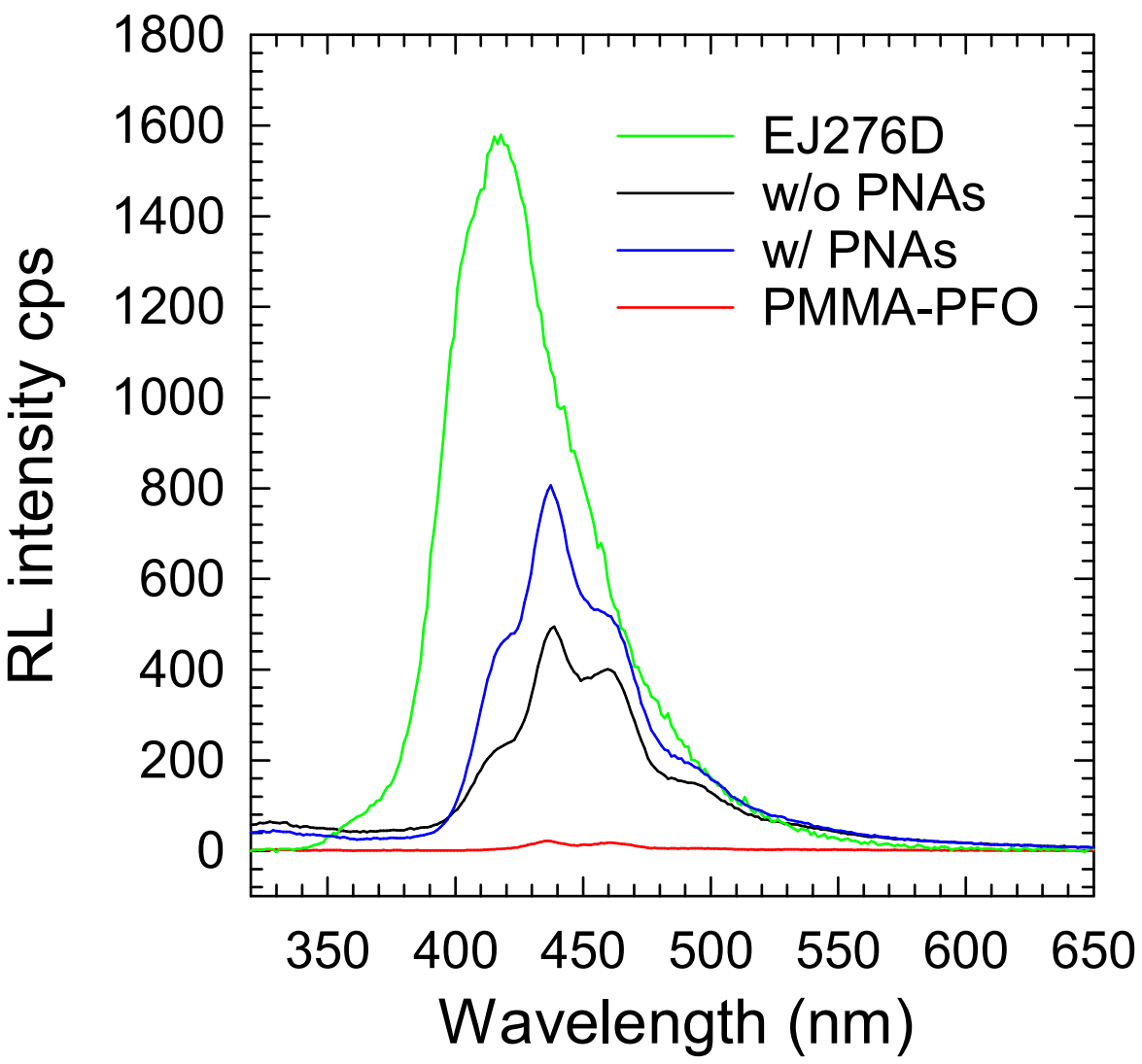

**Figure S8** Side by side RL spectra of the studied samples and commercial EJ276D scintillator specimen (thickness: 100μm) measured in identical excitation and collection conditions.

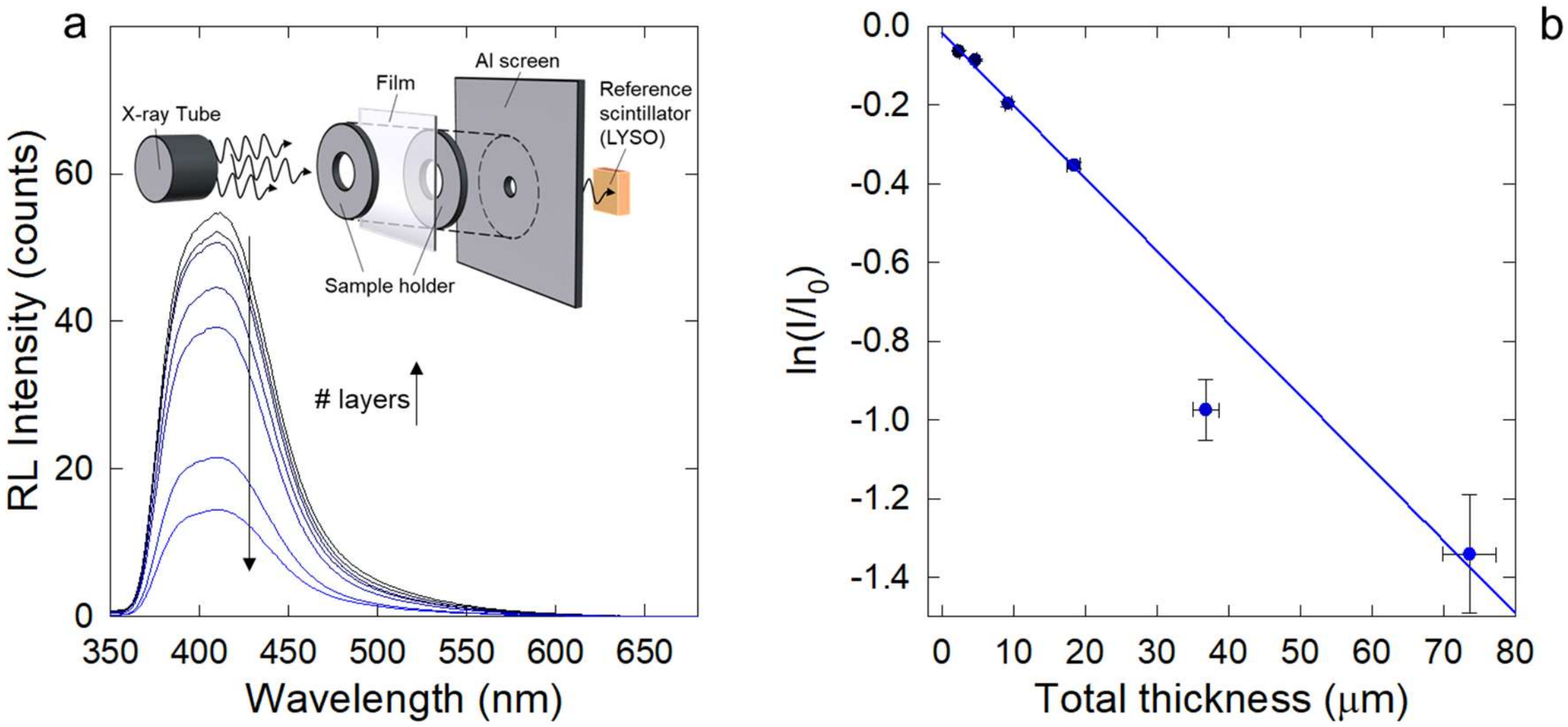

**Figure S9.** (a) RL spectra of a LYSO scintillator crystal collected while interposing a progressively increasing number of nanocomposite films containing PNAs, as illustrated in the schematic. (b) Natural logarithm of the ratio between the RL intensity measured with interposed films ($I$) and the RL intensity measured without the films in place ($I_0$), plotted as $ln(I/I_0)$. The solid line represents the linear fit, yielding an attenuation coefficient of 1.4 $cm^{-1}$, corresponding to an attenuation of 2.9±0.3% for a single 1.8 μm-thick layer.

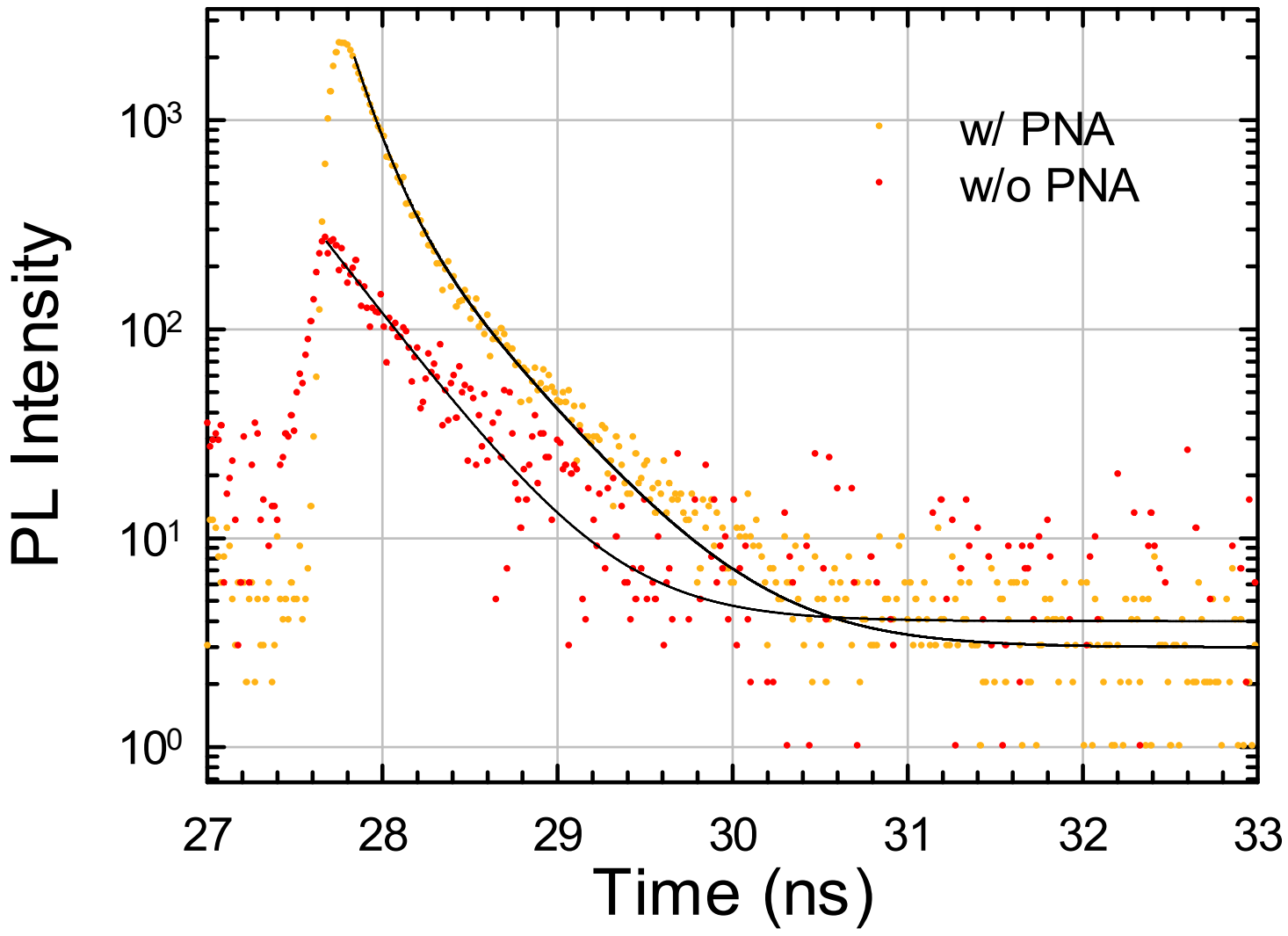

**Figure S10 Time-resolved photoluminescence spectra** of the studied sample set under 400nm excitation, measured under same configuration as the TRRL experiment. The extracted lifetimes are consistent with the corresponding radioluminescence decay times.

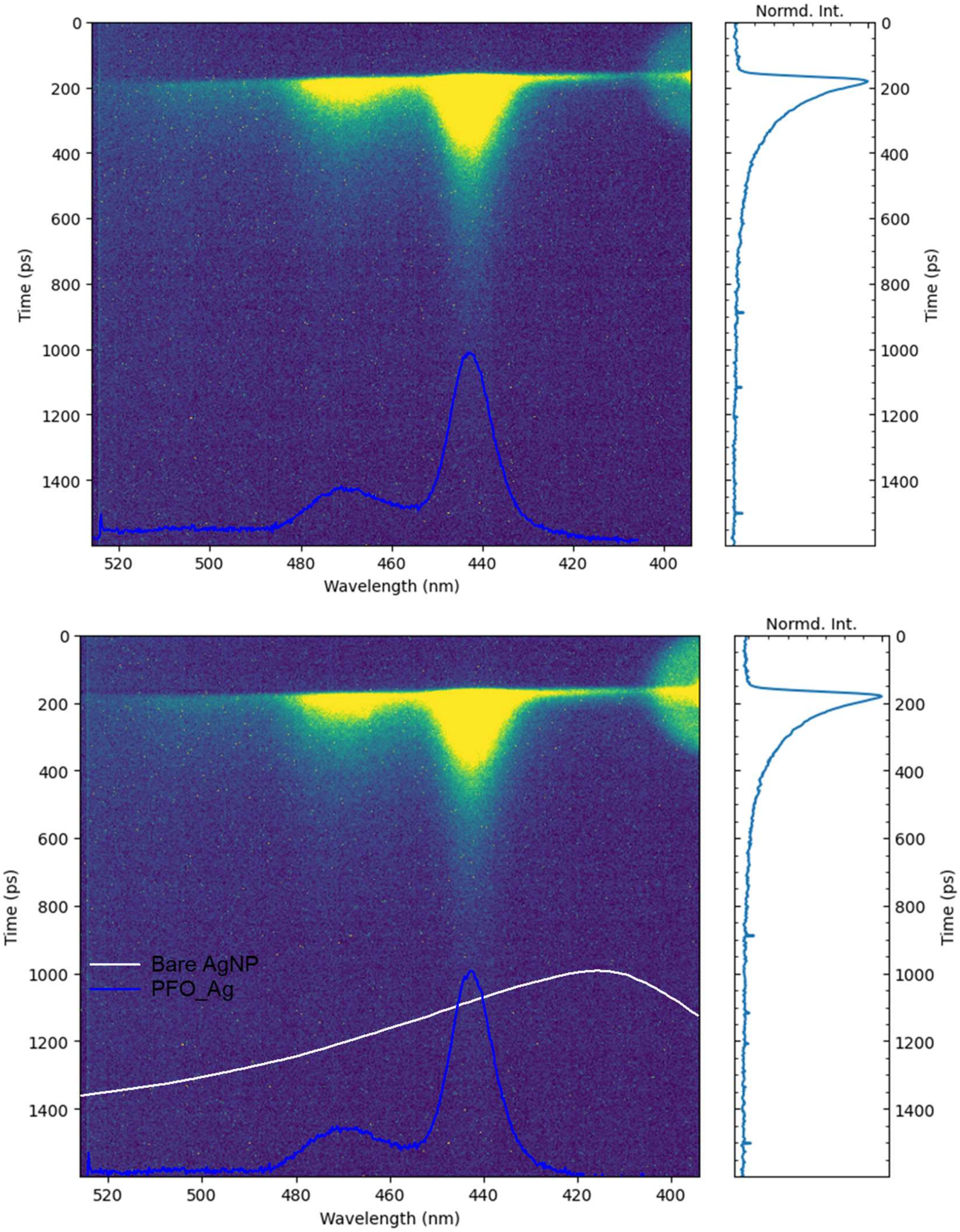

**Figure S11.** Off-resonant case (a) Streak camera measurements of PFO/PMMA composites overlaid with the photoluminescence (PL) and absorption spectra of Ag-$SiO_2$ PNA. Side panel: lifetimes extracted from the contour plot for PFO combined with thicker silica-shelled Ag-$SiO_2$ PNA, showing no change in spectral shape or lifetime.

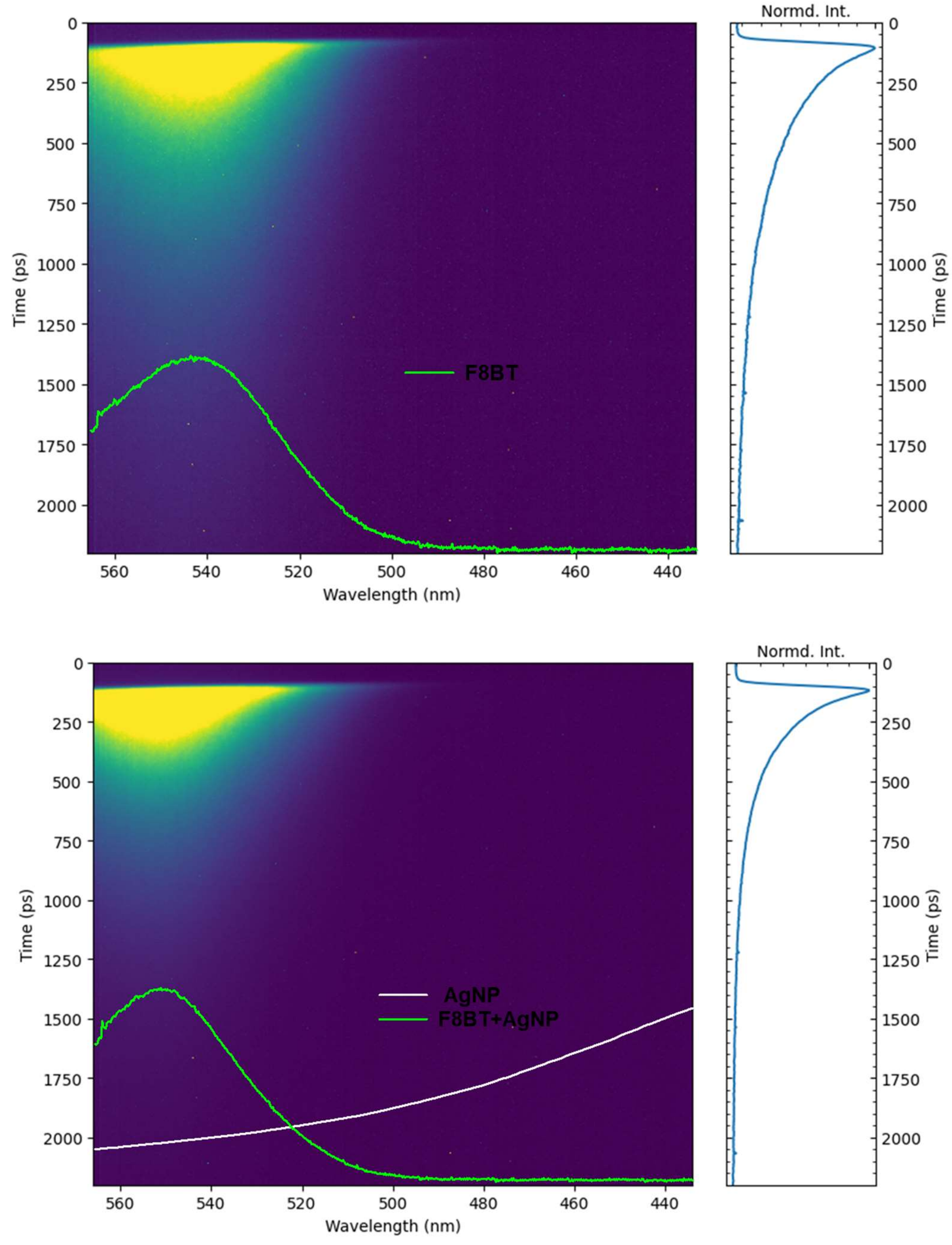

**Figure S12. Off-resonance case (b)** Streak camera measurements of F8BT overlaid with PL (F8BT) and absorption spectra of Ag-$SiO_2$ PNP, with side panel showing lifetimes extracted from the contour plot; emission lies in the off-resonance energy gap, and no change in spectral shape or lifetime is observed.

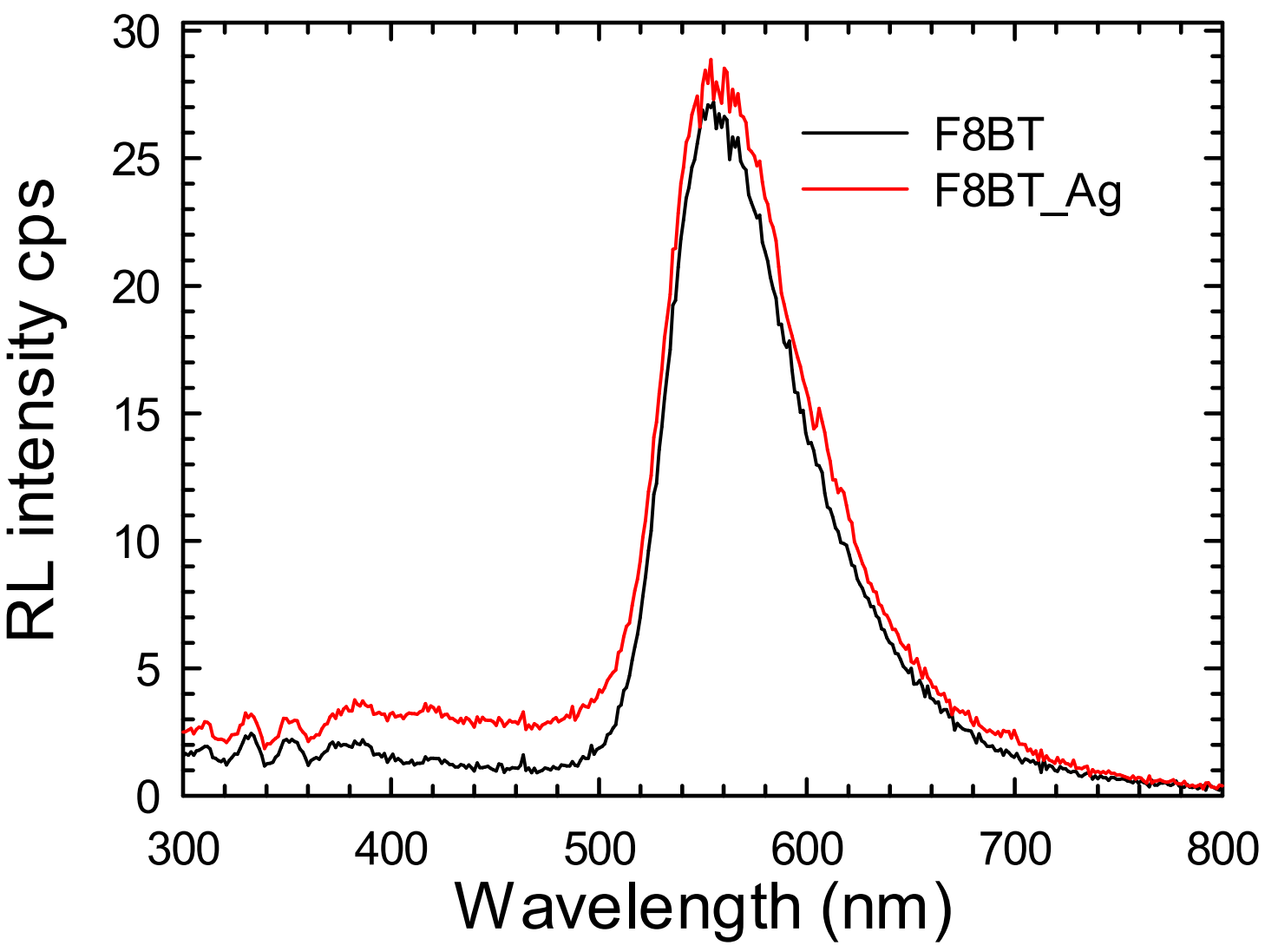

**Figure S13.** Radioluminescence spectra of F8BT/PMMA with the same amount of Ag-$SiO_2$ PNA, demonstrating that off-resonance RL intensity remains unchanged.

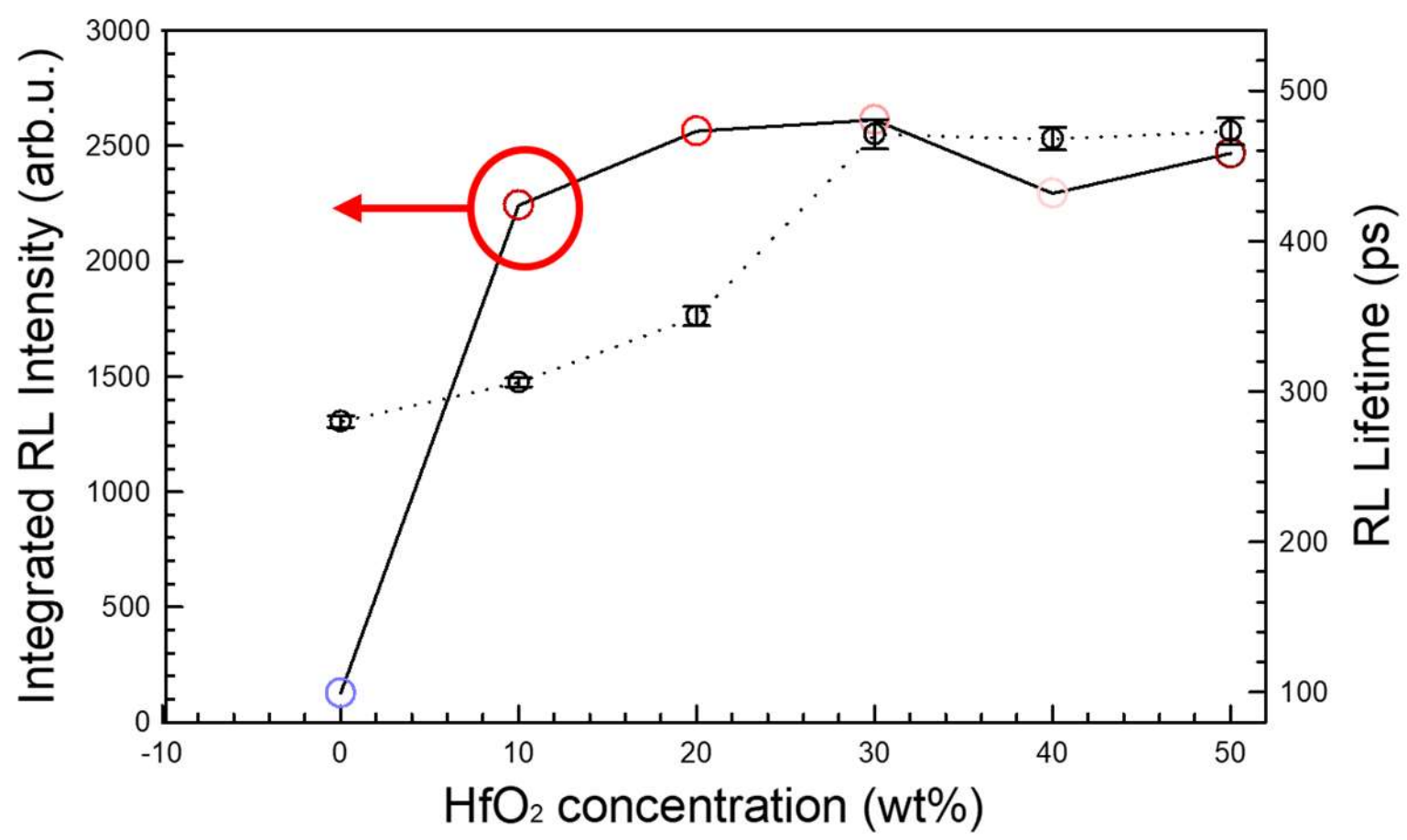

**Figure S14.** Radioluminescence intensity, and corresponding RL lifetime vs. $HfO_2$ content (10–50 wt %). Based on the provided supporting study, optimized PFO/PMMA/$HfO_2$ composites is chosen as PFO(1 wt%)/ $HfO_2$ (30 wt%)/PMMA, which showed ≈30× light-yield enhancement relative to non-sensitized systems.

**Supporting References**